\documentstyle[prb,aps,epsf]{revtex}
\include{psfig}

\begin{document}

\draft

%\wideabs{

\title{\bf Dynamics and phonon-induced decoherence of Andreev level qubit}
\author{ A.~Zazunov, V.~S. Shumeiko, and G. Wendin}
\address{Department of Microtechnology and Nanoscience,
Chalmers University of Technology, S-41296 G\"{o}teborg, Sweden}
\author{E.~N. Bratus'}
\address{B.\ Verkin Institute for Low Temperature Physics and
Engineering, 61103 Kharkov, Ukraine}
%\date{\today}

\maketitle

\begin{abstract}

We present detailed theory for Andreev level qubit, the system consisting of
a highly transmissive quantum point contact embedded in a superconducting
loop. The two-level Hamiltonian for Andreev levels interacting with quantum
phase fluctuations is derived by using the path integral method. We also derive
kinetic equation describing qubit decoherence due to interaction of the
Andreev levels with acoustic phonons. The collision terms are non-linear due
to fermionic nature of the Andreev states, leading to slow non-exponential
relaxation and dephasing of the qubit at temperature smaller than the qubit
level spacing.
\end{abstract}

%%%%%%%%%%%%%%%%%%%%%%%%%%%%%
\section{Introduction}

The possibility to employ Andreev bound levels in superconducting contacts
for quantum computation has been suggested in Refs.
\onlinecite{Lantz02,Zazunov03}. The proposed Andreev level qubit (ALQ)
consists of a highly transmissive, with reflectivity $R\ll 1$, quantum point
contact (QPC) embedded in a low inductance superconducting loop. In the ALQ,
quantum information is stored in the microscopic two-level system of Andreev
bound states in the contact. Hybridization of the clockwise and
counterclockwise persistent current states in the ALQ loop is produced by the
microscopic processes of electronic back scattering in the QPC. This is
different from the macroscopic superconducting flux
qubits\cite{Delft2000,Friedman00,Delft2003} and charge-phase
qubit\cite{Vion02}, where the hybridization is provided by charge
fluctuations on the tunnel junction capacitors. Thus requirement of the large
charging energy or large loop inductance is not critical for the ALQ. A
single ALQ consists of a pair of the Andreev bound levels belonging to the
same normal conducting mode in the QPC; a multimode QPC will form a qubit
cluster. The way of the ALQ operation is similar to the one of the
experimentally tested flux qubits\cite{Delft2000,Friedman00,Delft2003} - the
Andreev levels can be excited by driving biasing magnetic flux through the
qubit loop.\cite{Shumeiko93,Moriond,Lantz02} The read out method is also
similar to the flux and charge-phase qubits: the quantum state of the Andreev
levels determines the magnitude and direction of the persistent current
circulating in the loop, and also the magnitude of the induced flux.  Since
the quantum information is stored in the microscopic system, Andreev levels,
while the access for manipulation and readout is provided by macroscopic
persistent currents, ALQ occupies an intermediate place between the
microscopic solid state qubits (like localized spins on
impurities\cite{Kane98} or quantum dots\cite{Loss98}) and macroscopic
superconducting qubits.

During 90-s, the Josephson transport in superconducting QPCs has been
intensively investigated, and a number of remarkable experiments has been
performed on atomic size metallic QPCs  using controllable break junction
technique\cite{breakj:Post,breakj:Muller}, as well as on gated quantum
constrictions in 2D electron gas confined between superconductors.\cite{2DG}
The critical Josephson current and current-voltage characteristics have been
thoroughly examined in these experiments by applying current or voltage
bias.\cite{2DG,Scheer97,Ludoph00,Goffman00}  There was, however, one
experiment, particularly important in the qubit context, where the flux bias
has been implemented: Koops {\it et al.}\cite{Koops96} have inserted metallic
QPC in a SQUID, and evaluated the Josephson current-phase dependence by
measuring the induced flux with an inductively coupled SQUID magnetometer.
The measurements have been only reported for the equilibrium state.
Unfortunately, no experimental attempts to drive the QPC out of equilibrium
to some coherent or incoherent excited state have been performed so far.

The purpose of this paper is to present a theory for the Andreev level qubit
in more detail than that was possible in short communication.\cite{Zazunov03}
We will also consider the electron-phonon interaction as a potential
"intrinsic" source of the qubit decoherence, and derive a kinetic equation
for the qubit density matrix. The ALQ Hamiltonian and the kinetic equation
are derived by using a path integral method.
\cite{Ambegaokar,Eckern,Zaikin90} The
central technical difficulty here is to extend the method to contacts with
large transparency. This difficulty is overcome by incorporating exact
boundary condition in the QPC action. Another important point discussed in
the paper is the role of the charge electro-neutrality in the junction
electrodes, which affects the qubit Hamiltonian.

The fermionic nature of the Andreev levels does not affect the qubit
operation and qubit-qubit coupling but it plays important role for the qubit
decoherence. We find that the electron-phonon collision terms in the kinetic
equation for the ALQ differ qualitatively from the Bloch-Redfield master
equation\cite{Slichter} commonly applied to study decoherence of the
macroscopic superconducting qubits.\cite{Shnirman} This results in long
phonon-induced decoherence time for the ALQ, namely, at the temperature
smaller than the qubit level spacing, both the relaxation and dephasing
processes are governed by a power rather than exponential law.

The structure of the paper is the following. We discuss the model
description of the QPC in Section II, and then explain in detail, in
Section III, the path integral approach for transmissive QPC: we
consider the action for the contact and derive the effective
Hamiltonian for Andreev levels interacting with the quantum phase
fluctuations; the single-particle density matrix, and effective
current operator are also discussed in this section. In Section IV we
discuss averaging over fast phase fluctuations and derive an
effective Hamiltonian for the qubit, this procedure is extended in
Section V to the two inductively coupled qubits to derive direct
qubit-qubit interaction. Section VI is devoted to the electron-phonon
interaction: we derive effective action for the Andreev level-phonon
interaction and derive corresponding collision terms in the equation
for the qubit density matrix; then we present solution of the kinetic
equation, and evaluate the decoherence rate.

%%%%%%%%%%%%%%%%%%%%%%%%%%%%%%%%%%%%%%%%%%%%%%%%%%%%%%%%%%%%%%%%%%%%%%%%%%%%

\section{Contact Hamiltonian}

Let us consider a superconducting quantum point contact with bulk 3D
electrodes. We model the contact with a smooth on the atomic scale
(adiabatic) constriction and assume a local scatterer
situated in the neck of constriction (see Fig. 1)
producing weak electronic back scattering with the reflectivity $R \ll 1$.
We further assume that the constriction supports a single conducting mode.
%%%%%%%%%%%%%%%%%%%%%%%%%%%%%%%%%%%%
\begin{figure}[h]
\centerline{\psfig{figure=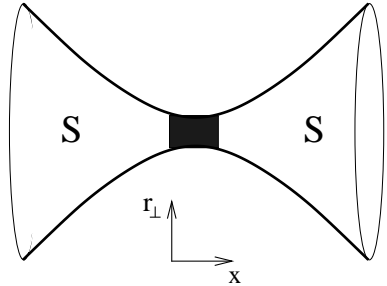,width=5cm}}
\label{fig:constriction}
\vspace{.5cm} \caption { Adiabatic superconducting constriction with a local
scatterer (dark region) in the neck. The length of the constriction is small
on the scale of the superconducting coherence length but large on the atomic
scale. }
\end{figure}
%%%%%%%%%%%%%%%%%%%%%%%%%%%%%%%%%%%%%%
We adopt the mean field approximation for the
electrons in the contact, which are described with the Hamiltonian,
\begin{equation}
 H_{e} = \int d{\bf r} \;\Psi^\dagger({\bf r},t) h
({\bf
r},t)
\Psi({\bf r},t) + {1\over 2} \, C V^2(t),
\label{He}
\end{equation}
where the first term is the BCS Hamiltonian for the bulk superconducting
electrons, $\Psi({\bf r},t)$ being the two-component Nambu field operator,
and the second term describes the charging energy of the contact capacitor
$C$ \cite{Ambegaokar,Eckern,Zaikin90}.
The single-particle Hamiltonian $h$ in Eq.
(\ref{He}) has the form,
\begin{equation}
h = \left[ \,
\frac{(- i \hbar {\bf \nabla} - (e/c) {\bf A}({\bf
r},t) \,
\sigma_z
)^2}{2 m}- \mu + U({\bf r})
+ e \varphi({\bf r},t) \, \right] \sigma_z +
\Delta({\bf r},t) \, e^{i \sigma_z \chi({\bf r},t)} \sigma_x
\,,
\label{h}
\end{equation}
where $\Delta({\bf r},t)$ and $\chi({\bf r},t)$ are, respectively,
the modulus and phase of the superconducting order parameter,
the potential $U({\bf r})$ accounts
for the confinement of electrons within the contact as well as the electron
scattering, while $\varphi({\bf r},t)$ and ${\bf A}({\bf r},t)$ are
electromagnetic potentials. The voltage drop at the contact, $V(t)$, in Eq.
(\ref{He}) is related to the discontinuity of the electric potential at the
contact, $V(t) = \varphi(-0,t) - \varphi(+0,t)$. To investigate the
decoherence effects, we allow the electrons in the electrodes to interact
with acoustic phonons, and include corresponding electron-phonon interaction
and phonon terms in the total Hamiltonian of the contact,
\begin{equation}
 H_{c} =  H_{e} +  H_{e-ph} +  H_{ph} \,.
\label{contactH}
\end{equation}

It is convenient to introduce quasiclassical (Andreev) approximation for the
superconducting electrons. Following standard procedure, we eliminate rapidly
varying in space potential, $U({\bf r})$, by introducing quasiclassical wave
functions of the single conducting mode in the left and right electrodes,
\begin{equation}
\Psi({\bf r},t) = \sum_{\sigma = \pm} \psi_\perp({\bf r}_\perp,x)
\, e^{i \sigma \int dx \, k(x) } \, \psi^{(\sigma)}(x,t) \,,
\label{Psi}
\end{equation}
and couple these wave functions by means of a normal electron scattering
matrix. In Eq. (\ref{Psi}), $\psi^{(\sigma)}(x)$ are slowly varying 1D
envelopes for the longitudinal electron motion ($\sigma = \pm$ indicates the
direction of the motion), $\psi_\perp({\bf r}_\perp,x)$ is a normalized wave
function of the transverse motion with energy $E_\perp(x)$, $p(x) = \hbar
k(x) = \sqrt{2m \left( \mu - E_\perp(x) \right)} = mv(x)$ is the
quasiclassical longitudinal electronic momentum. The coupling of the
quasiclassical envelopes in the left ($L$) and right ($R$) electrodes,
$\psi_{L,R}$, is conveniently described by the transfer matrix,
\cite{Shumeiko93,Shumeiko97}
\begin{equation}
%\sqrt{D} \, e^{- i \sigma_z \phi(t)/2} \,
\left(
\begin{array}{c}
\psi^{(+)}_L\\ \psi^{(-)}_L
\end{array}
\right) (0,t)  = \hat T
\left(
\begin{array}{c}
\psi^{(+)}_R \\ \psi^{(-)}_R
\end{array}
\right) (0,t)\;,
\label{BC}
\end{equation}
\begin{equation}
\hat T =
\left(
\begin{array}{cc}
1/d & r^\ast/d^\ast \\ r/d & 1/d^\ast
\end{array}
\right) \,.
%%\label{}
\end{equation}
Here $d$ and $r$ are the energy independent transmission and reflection
amplitudes, respectively. Since any observable quantity is expressed through
a bilinear combination of the envelopes with the same $\sigma$, the energy
independent scattering phases can be eliminated from the boundary condition
(\ref{BC}), hence, without loss of generality, the scattering amplitudes will
be further assumed to be real, $d=\sqrt D$, $r=\sqrt R$, where $D$ and $R$
are the transmission and reflection coefficients of the contact,
respectively.

The electromagnetic potentials, $\varphi({\bf r},t)$, ${\bf A}({\bf r},t)$,
and the complex order parameter in Eq. (\ref{h}) are to be found from the
Maxwell equations and the self-consistency equation. It is convenient to
present the Hamiltonian in a gauge invariant form by extracting the phase of
the order parameter using a local gauge transformation, $\Psi({\bf r},t)
\rightarrow e^{i \sigma_z \chi({\bf r},t)/2} \, \Psi({\bf r},t) \,.$ Then the
superfluid momentum ${\bf p}_s = \hbar {\nabla} \chi/2 -(e/c) {\bf A}$, and
the gauge-invariant electric potential $\tilde\varphi = \hbar \dot{\chi}/2  +
e\varphi$, appear in the quasiclassical Hamiltonian of the electrode,
\begin{equation}
h^{(\sigma)} = \sigma v (-i\hbar\partial_x)\,\sigma_z + \sigma v p_s +
\tilde\varphi \, \sigma_z + \Delta \, \sigma_x,
\label{Hinv}
\end{equation}
while the phase difference
$\phi(t) = \chi_R(0,t) - \chi_L(0,t)$,
appears in the boundary condition,
\begin{equation}
\hat T \rightarrow e^{i\sigma_z\phi(t)/2} \; \hat T \,.
\label{Tphi}
\end{equation}
In the bulk metallic electrodes with good screening, and at small frequencies
relevant for the problem, the gauge-invariant fields,
$\tilde\varphi({\bf r},t)$, ${\bf p}_s({\bf r},t)$,
are to be found from the electro-neutrality
condition, and the current conservation,\cite{Galaiko,Artemenko,Aronov}
\begin{equation}
\delta n({\bf r},t)=0,\;\;\; \nabla{\bf j}({\bf r},t)=0 ,
 \label{neutr}
\end{equation}
where $n({\bf r},t)$ is the electronic density.
In the electrodes, the charge
imbalance relaxation yields the equilibrium relation, $\tilde\varphi =
(\partial n/\partial \mu)^{-1} \delta n $, on the distance exceeding the
electric field penetration length. Furthermore, in the absence of normal
dissipative current, ${\bf p}_s$ is proportional to the total current
density, ${\bf j} = (en/m){\bf p}_s $, which is negligibly small far from the
contact due to rapid spreading out of the current in the point contact
geometry. Thus the conditions (\ref{neutr}) yield complete cancellation of
the electromagnetic potentials in the electrodes,
\begin{equation}
\tilde\varphi(x,t)=0,\;\;\;  p_s(x,t)=0 \,.
\label{neutr1}
\end{equation}
Taking  into account that the modulus of the order parameter far from the
contact is equal to the equilibrium value,\cite{KulikOm}
$\Delta = {\rm const}$,
we conclude that the Hamiltonian (\ref{Hinv}) approaches the equilibrium form.

Potential $\tilde\varphi$ can be expressed through the electric field $E$,
$\tilde\varphi=\hbar\dot{\bar\chi}/2 + e \int dx \, E$, by introducing the
gauge invariant phase $\bar\chi = \chi - (2e/c\hbar)\int dx \, A =
(2/\hbar)\int dx \, p_s$. The spatial distribution of these quantities is
illustrated on Fig. 2. Since $\tilde\varphi$ vanishes far from the contact,
and $\bar\chi$ is constant according to Eq. (\ref{neutr1}), then the
gauge-invariant phase difference across the contact
$\dot{\bar\phi} =
\dot{\bar\chi}_R (0,t)- \dot{\bar\chi}_L (0,t)$ is related to the voltage
drop, $V = \int_{-\infty}^\infty dx \, E$ ,
\begin{equation}
\dot{\bar\phi}(t) = {2eV\over \hbar}
\label{Jos}
\end{equation}
(Josephson relation). In the SQUID, this relation is equivalent to the phase
versus flux relation, $\bar\phi = 2e \Phi/\hbar c$, since the voltage drop
across the contact is generated by time variation of the magnetic flux $\Phi$
threading the SQUID. Finally, we notice that the gauge-invariant phase
difference, $\bar\phi$, rather than $\phi$ enters the boundary condition
(\ref{Tphi}), which can be explicitly seen by extracting the Aharonov-Bohm
phase from the transfer matrix. Thus the gauge-invariant phase difference
remains the only free collective variable whose dynamics is determined by the
electrodynamic environment of the contact. Below we will not distinguish
between $\bar\phi$ and $\phi$, because the difference is negligibly small in
the QPC.

%%%%%%%%%%%%%%%%%%%%%%%%%%%%%%%%%%%%%%%
\begin{figure}[h]
\centerline{\psfig{figure=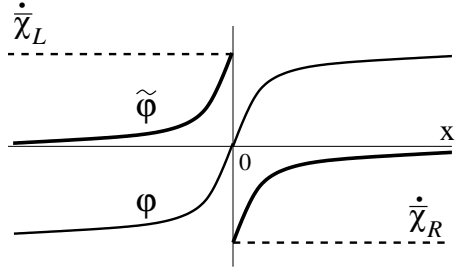,width=6cm}} \label{fig:figPhi}
\vspace{.5cm} \caption {Spatial distribution of the electric potential,
$\varphi$ (thin line), gauge invariant potential, $\tilde\varphi$ (bold
line), and the time derivative of the gauge invariant phase,
$\dot{\bar\chi}$ (dashed line), in the vicinity of the contact. }
\end{figure}
%%%%%%%%%%%%%%%%%%%%%%%%%%%%%%%%%%%%%%%%%%%

Proceeding with discussion of the interaction of electrons with phonons, we
consider only longitudinal acoustic phonons and describe the interaction
within the deformation potential approximation,
\begin{equation}
 H_{e-ph} = \gamma \int d{\bf r} \,
\nabla{\bf u}({\bf r},t) \,
\Psi^\dagger({\bf r},t) \sigma_z \Psi({\bf r},t),
\label{Heph}
\end{equation}
where $\gamma$ is the deformation potential
constant, ${\bf u}({\bf r},t)$ is the phonon field operator,
\begin{equation}
{\bf u}({\bf r})= \sum_{\bf q}
\sqrt{{\hbar\over 2 \rho V \Omega_q}}\,
{{\bf q}\over q}\, \left( b_{\bf q} \, e^{i{\bf qr}} +
b^\dagger_{\bf q} \, e^{-i{\bf qr}}  \right)
\;,\;\;\;
\Omega_q = sq \,,
%%\label{}
\end{equation}
$s$ is the sound velocity, and $\rho$ is the crystal mass density.
The Hamiltonian of free phonons has a standard form,
\begin{equation}
H_{ph}= \sum_{\bf q} \hbar\Omega_q
\left( b^\dagger_{\bf q} b_{\bf q} + 1/2 \right).
%%\label{}
\end{equation}

Our strategy now will be to derive effective Hamiltonian for the Andreev
levels including interaction with phonons. If the phase difference would be a
classical variable, this derivation can be in principle done by direct
truncation of the Hamiltonian (\ref{contactH}). However, in the presence of
quantum phase fluctuations it is convenient to apply the path integral
technique.

 % % % % % % % % % % % % % % % % % % % % % % % % % % % % % % % % % % % % % % %
 % %
\section{Contact effective action}

Let us consider the whole system,
the QPC and superconducting loop (see Fig. 3),
and introduce the path integral representation for the propagator,
\begin{equation}
{\cal U} = \int {\cal D}^2 \psi_L {\cal D}^2 \psi_R \, {\cal D} \{ X_{\bf q}
\} \, {\cal D} \phi \, e^{i \int dt L_{tot} / \hbar} \;,\;\;\; X_{\bf q} =
b_{\bf q} + b_{\bf q}^\ast \,. \label{U}
\end{equation}
The Lagrangian of the system $L_{tot}$ consists of the contact part $L_c$,
and the part describing the circulating current in the loop. The latter is
conveniently combined with the charge term from the electronic Hamiltonian
(\ref{He}) giving the Lagrangian of the loop oscillator $L_{osc}$,
\begin{equation}
L_{tot} = L_c + L_{osc}, \;\; L_{osc}= \left({\hbar \over 2e}\right)^2 \left(
{C\over 2} \, (\partial_t\phi)^2 - {c^2 \over 2L} \, (\phi - \phi_e)^2\right)
\,, \label{Losc}
\end{equation}
where $\phi_e$ corresponds to the bias magnetic flux, and $L$ is the loop
inductance. The remaining part of the contact Lagrangian consists, similar to
Eq. (\ref{contactH}), of the electronic part, phonon part, and
electron-phonon interaction,
\begin{equation}
L_c = L_e +  L_{ph} + L_{e-ph} \,.
\label{contactL}
\end{equation}

In the quasiclassical approximation,
the electronic Lagrangian splits into two parts, $L_\alpha$, $\alpha = L,R$,
corresponding to the left and right electrodes,
\begin{equation}
L_\alpha = \sum_{\sigma = \pm} \int dx \,
\bar{\psi}_\alpha^{(\sigma)} (x,t) \,
{\cal L}^{(\sigma)}(x,t) \,
\psi_\alpha^{(\sigma)}(x,t) \,, \;\;
{\cal L}^{(\sigma)}(x,t) = i \hbar \partial_t +
\sigma v i \hbar \partial_x \sigma_z - \Delta  \sigma_x \,,
%%\label{}
\end{equation}
and the third part $L_{BC}$ discussed in detailed in the next section,
which accounts for the boundary condition. Noting that relaxation processes
are caused by phonons with small wave vectors
compared to the Fermi wave vector, $q\ll k$,
transitions between the states $\psi^{(+)}$ and $\psi^{(-)}$
are forbidden, and the electron-phonon Lagrangian can be written on the form,
\begin{equation}
 L_{e-ph} = - \gamma
\sum_{\alpha = L,R} \sum_{\sigma = \pm}
\int d {\bf r} \; |\psi_\perp|^2 \,
\bar{\psi}_\alpha^{(\sigma)} (x,t) \, \sigma_z \,
\psi_\alpha^{(\sigma)}(x,t) \nabla {\bf u}({\bf r},t)\,.
\label{Le-ph}
\end{equation}
The phonon Lagrangian is given by equation,
\begin{equation}
L_{ph} = {1 \over 2} \sum_{\bf q} \, (\hbar / 2 \Omega_q) \, X_{\bf q}
\left[( i \partial_t)^2 - \Omega_q^2 \right] X_{\bf q}
\,.
\label{Lph}
\end{equation}
 %
%%%%%%%%%%%%%%%%%%%%%%%%%%%%%%%%%
\begin{figure}[h]
\centerline{\psfig{figure=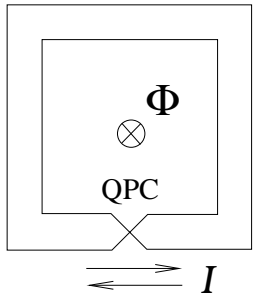,width=3.3cm}} \label{fig:squid}
\vspace{.5cm}
\caption{ Sketch of the Andreev level qubit: a low inductance
superconducting loop with a quantum point contact (QPC). $\Phi$ is the
magnetic flux; the arrows indicate fluctuating persistent currents. }
\end{figure}
%%%%%%%%%%%%%%%%%%%%%%%%%%

 % % % % % % % % % % % % % % % % % % % % % % % % % % % % % % % % % % % % % % %
%
% %
\subsection{Boundary condition}

The boundary condition (\ref{BC}) is valid for any contact transparency. To
include this boundary condition in the path integral formulation, we
introduce an additional term in the Lagrangian\cite{Zazunov03},
\begin{equation}
L_{BC}=
\bar{\eta}(t) \sum_\sigma \left[
\sqrt{D} \, e^{-i \sigma_z \phi(t)/4}
\psi_L^{(\sigma)}(0,t)
-
(1 + \sqrt{R}) e^{i \sigma_z \phi(t)/4}
\psi_R^{(\sigma)}(0,t) \right]
+ h.c. \, ,
\label{LBC}
\end{equation}
where $\eta$ is an auxiliary fermionic Nambu field playing the role of
a Lagrange multiplier. Correspondingly, the integration over $\eta$ is
to be included in the propagator in Eq. (\ref{U}), giving the following
form for the electronic part of the propagator,
\begin{equation}
{\cal U}_e = \int {\cal D}^2 \eta \, {\cal D}^2 \psi_L
{\cal D}^2 \psi_R \, e^{i\int dt (L_L + L_R + L_{BC}) /\hbar} \,.
\label{Ue}
\end{equation}
Let us prove that such a Lagrangian indeed generates the boundary
condition (\ref{BC}), (\ref{Tphi}). To this end, it is convenient
to make a step back and restore a non-quasiclassical form for the fermionic
field,
\begin{equation}
\psi(x,t) = \sum_\sigma e^{i\sigma \int
kdx}\,\psi^{(\sigma)}(x,t) \,,
%%\label{}
\end{equation}
in the bulk part of the Lagrangian,
\begin{equation}
L_L \,+ \, L_R = \int_{-\infty}^{\infty} dx \,
\bar\psi(x,t) {\cal L}(x,t) \psi(x,t),
\;\;\;
{\cal L}(x,t) = i\hbar\partial_t -
[(- \hbar^2/2m)\partial_x^2 + E_\perp - \mu ] \, \sigma_z - \Delta \, \sigma_x
\,.
%%\label{}
\end{equation}
The dynamic equations and the boundary condition
result from the zero variation of the action
$S_e = \int dt \, ( L_L + L_R + L_{BC})$
with respect to $\bar{\psi}_{L,R}$ and $\bar{\eta}$,
\begin{equation}
{\delta S_e\over \delta \bar\psi_{L,R}(x,t)} =0 \;,\;\;\; {\delta S_e\over
\delta \bar\eta(t)} = 0 \,,
%%\label{}
\end{equation}
or in the explicit form,
\begin{eqnarray}
{{\cal L}}(x,t) \, \psi_L (x,t) & = &
- \sqrt{D}\, \delta(x) \, e^{i \sigma_z \phi(t)/4}
\,
\eta(t)
\;,\nonumber \\
 %\end{equation}
 %\begin{equation}
{{\cal L}}(x,t) \, \psi_R (x,t)  & = & (1 +
\sqrt{R}) \,
\delta(x) \,
e^{-i \sigma_z \phi(t)/4} \,\eta(t) \,,
\label{eom1+2}
\end{eqnarray}
and
\begin{equation}
\sqrt{D} \, e^{-i \sigma_z \phi(t)/4} \,
\psi_L(0,t) =
(1 + \sqrt{R}) \, e^{i \sigma_z \phi(t)/4} \,
\psi_R(0,t)
\,.
\label{eom3}
\end{equation}
Integrating Eqs. (\ref{eom1+2}) over $x$
in a small vicinity of $x = 0$ yields the relations,
\begin{eqnarray}
(\hbar^2/2m) \, \sigma_z \partial_x \psi_L(0,t)
& = & \sqrt{D} \, e^{i \sigma_z \phi(t)/4} \eta(t)
\;, \nonumber\\
(\hbar^2/2m) \, \sigma_z \partial_x \psi_R(0,t)
& = & (1 + \sqrt{R}) \, e^{-i \sigma_z \phi(t)/4} \,
\eta(t) \,.
\label{eom1+2a}
\end{eqnarray}
Then introducing again the quasiclassical envelopes and
combining Eqs. (\ref{eom3}), (\ref{eom1+2a}) with the quasiclassical relation,
\begin{equation}
\partial_x \psi_{L,R}(0,t) = i k
\sum_{\sigma=\pm} \sigma \,
\psi_{L,R}^{(\sigma)}(0,t) \,,
 %%\label{}
\end{equation}
we get the boundary condition equivalent to Eqs.
(\ref{BC}), (\ref{Tphi}).

 % % % % % % % % % % % % % % % % % % % % % % % % % % % % % % % % % % % % % % %
%
% % %
\subsection{Effective action for Andreev levels}

Now we are prepared to derive an effective action for the
Andreev levels. Following the procedure of Ref.
\onlinecite{Ambegaokar}, we integrate out fast electronic fields
$\psi_{\alpha}$ in Eq. (\ref{Ue}),
\begin{equation}
e^{ i S_\eta ^0/\hbar} = \int
\prod_{\alpha = L,R} \prod_{\sigma = \pm}
{\cal D} \bar{\psi}_\alpha^{(\sigma)} \,
{\cal D} \psi_\alpha^{(\sigma)} \,
\exp \left\{ \frac{i}{\hbar}
\int dt \, L_e \right\} \,.
 \label{intpsi}
\end{equation}
Then we are left with the effective action
$S_\eta^0$,
which
contains only variables, $\eta(t)$ and $\phi(t)$,
\begin{equation}
S_\eta^0 = - \int dt_1 dt_2 \, \bar{\eta}(t_1)
\left[
D e^{-i \sigma_z \phi(t_1)/4} g(t_1 - t_2) \, e^{i
\sigma_z
\phi(t_2)/4} +
(1+\sqrt{R})^2 \, e^{i \sigma_z \phi(t_1)/4} \,
g(t_1 -
t_2)
\,
e^{-i \sigma_z \phi(t_2)/4}\right] \eta(t_2) \,,
\label{Seta0}
\end{equation}
where $g(t)$ is given by the Fourier component,
\begin{equation}
g_\omega = - {\hbar \omega + \Delta \, \sigma_x \over \hbar v
\sqrt{\Delta^2 - (\hbar \omega)^2}}\,.
\label{contact:g}
\end{equation}
Connection between the effective action (\ref{Seta0}) and the Andreev levels
can be established by considering the case of time independent phase,
$\phi=const$. Indeed, by writing the effective action in the Fourier
representation,
\begin{equation}
S_\eta^0 = \frac{2 (1 + \sqrt{R})}{\hbar v} \int \frac{d
\omega}{\sqrt{\Delta^2 - (\hbar \omega)^2}} \, \bar{\eta}_\omega \left[ \hbar
\omega + \Delta \left( \cos \frac{\phi}{2} \, \sigma_x - \sqrt{R} \, \sin
\frac{\phi}{2} \, \sigma_y \right) \right] \eta_\omega \,.
\label{Seta0_Fourier}
\end{equation}
we identify the spectrum of the system, $\hbar \omega = \pm E_a(\phi)$, by
calculating eigenvalues of the matrix in the brackets,
\begin{equation}
E_a(\phi) = \Delta \sqrt{\cos^2(\phi/2) + R \sin^2 (\phi/2)}\,;
\label{Ea}
\end{equation}
this equation coincides with well known Andreev level
spectrum\cite{Furusaki90,Beenakker91} (see Fig. 4). Thus we conclude that the
fermionic field $\eta$ represents the Andreev levels.

Proceeding to a time-dependent $\phi(t)$, we restrict its time variation
speed to small values, $\hbar \partial_t \phi/4 \ll \Delta$. Furthermore, the
dynamics of the Andreev levels is also to be slow, $E(\phi) \ll \Delta$,
which implies that the contact reflectivity must be small, $R\ll 1$, ALQ must
be biased at $\phi_e \approx \pi$, and the amplitude of the quantum phase
fluctuations $\tilde\phi(t) = \phi(t)-\phi_e$, must be sufficiently small,
$\tilde\phi \ll \phi_e$. We emphasize that the constraint on the amplitude of
the phase fluctuations is actually provided in our case due to the loop
geometry of the electrodes having sufficiently small inductance; the
constraint is important to suppress the Landau-Zener transitions to the
continuum states. Under the imposed conditions, the non-local in time kernel
$g(\omega)$, Eq. (\ref{contact:g}), can be replaced by a constant value
(adiabatic approximation), $\sqrt{\Delta^2 - (\hbar \omega)^2} \rightarrow
\sqrt{\Delta^2  - E_a^2} = \zeta_e$ with $E_a = E(\phi_e)$, leading to the
equation,
\begin{equation}
S_\eta^0 = \frac{2 (1 + \sqrt{R})}{\hbar v\zeta_e}
\int dt
\,
\bar{\eta}(t) \left[
i \hbar \partial_t + \frac{\sqrt{R}}{4} \, \hbar
\dot{\phi}
\, \sigma_z
+
\Delta \left(
\cos \frac{\phi}{2} \, \sigma_x - \sqrt{R} \, \sin
\frac{\phi}{2} \,
\sigma_y
\right)
\right] \eta(t) \,.
\label{Seta0_slow}
\end{equation}
We eliminate the term with the phase time derivative
in Eq. (\ref{Seta0_slow}) by transforming $\eta$,
\begin{equation}
\eta \rightarrow
\left( \frac{\hbar v \, \zeta_e}{2(1 + \sqrt{R})}
\right)^{1/2}
e^{i \sigma_z \sqrt{R} \phi/4} \,
e^{i \sigma_y \pi/4} \, \eta \,.
\label{transformation}
\end{equation}
Then we finally arrive at the effective action,
\begin{equation}
S_\eta^{0} = \int dt \, \bar{\eta}(t)
\left[ \, i \hbar \partial_t - h_a \right] \eta(t) \,,
\label{Seta0_final}
\end{equation}
where
\begin{equation}
h_a = \Delta \, e^{-i \sigma_x \sqrt{R} \, \phi /2} \left( \cos
\frac{\phi}{2} \, \sigma_z + \sqrt{R} \, \sin \frac{\phi}{2} \, \sigma_y
\right) 
\label{Ha}
\end{equation}
presents the effective single-particle Hamiltonian for the two-level Andreev
system.\cite{Zazunov03}
The Hamiltonian in Eq. (\ref{Ha}) differs by the exponential
pre-factor from the two-level Hamiltonian derived in
Refs.\onlinecite{Ivanov99,Averin} and further employed in
Refs.\onlinecite{IvanovFeig:ph,LYeyati01}. This factor appears in the present
derivation after the electric potential has been included into consideration
in Eq. (\ref{He}) to provide the electro-neutrality in the electrodes [see
text after Eq. (\ref{Tphi})]. Both Hamiltonians are equivalent under
stationary conditions, $\partial_t\phi = 0$, and the difference is not
important for the adiabatic dynamics. However, in general, the pre-factor is
important, e.g. for derivation of correct equation for the current operator
in Eq. (\ref{calI}).

%%%%%%%%%%%%%%%%%%%%%%%%%%
\begin{figure}[h]
\centerline{\psfig{figure=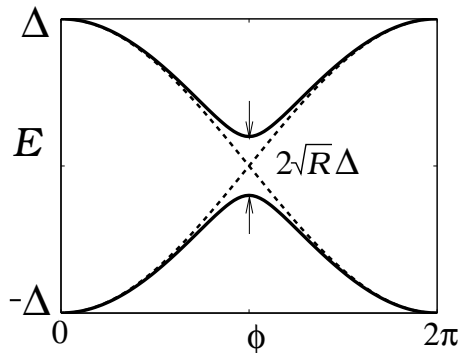,width=6cm}} \vspace{.5cm} \caption{
Spectrum of the Andreev levels in a QPC with finite reflectivity ($R=0.04$)
(solid line), the level anticrossing is produced by electronic back
scattering; at $R=0$ the Andreev levels (dashed line) coincide with the
current eigen states. }
\end{figure}
%%%%%%%%%%%%%%%%%%%%%%%%%%%%%%%%%%%%

It is instructive to compare the case of the transparent
contact considered here with the case of a tunnel contact extensively studied
in the MQC theory.\cite{Ambegaokar,Eckern,Zaikin90}
Physical difference between the two cases is that
the Andreev level system in transparent contacts is a slow one,
while in tunnel
contacts it is a fast one because the Andreev level energy
in tunnel contacts is close to $\Delta$. Within the present
formalism, the integration over $\psi$ fields is similar in both cases.
However, the next step, the adiabatic approximation in Eqs. (\ref{Seta0}),
(\ref{contact:g}) is not allowed in the tunnel limit;
instead one should perform also the integration over $\eta$ in
Eq. (\ref{Ue}), and make an expansion over small $D$.
The result of this calculation, presented in Appendix A,
coincides with the results of Refs.
\onlinecite{Ambegaokar,Eckern,LarkinOvch83}.

%%%%%%%%%%%%%%%%%%%%%%%%%%%%%%%%%%%%%%%%

\subsection{Andreev levels density matrix  }

The Hamiltonian (\ref{Ha}) governs the free evolution of the Andreev
variable $\eta$,
\begin{equation}
\label{equation_eta}
{\delta S_\eta^0\over \delta\bar\eta(t)} = 0\,,\;\;\;
 i\hbar\partial_t \eta = h_a\eta.
\end{equation}
The same Hamiltonian governs the free evolution of the {\em single-particle}
density matrix of the Andreev levels $\rho_a(t)$ defined via statistical
average,
\begin{equation}
\rho_{a}(t) = \langle \hat\eta(t) \hat\eta^\dagger(t) \rangle,
\label{rhoa}
\end{equation}
where $\hat\eta$ denotes the fermionic operator corresponding to the
Grassmann field $\eta$, and the average is taken over all electronic states.
The density matrix $\rho_a$ is a $2\times 2$ matrix in the Nambu space, which
satisfies the normalization condition Tr$\,\rho_a=1$. The statistical average
in Eq. (\ref{rhoa}) is represented, after the averaging over $\psi_{L,R}$, by
a path integral,
\begin{equation}
\rho_a(t) = \int {\cal D}^2 \eta \,\eta(t)\bar\eta(t)
\,e^{iS^0_\eta/\hbar} \,.
\label{rhoaPath}
\end{equation}
Then calculating the time derivative, using Eq. (\ref{equation_eta})
and the conjugated equation, we get,
\begin{equation}
i\hbar \partial_t\rho_a = [h_a \,, \rho_a]. 
\label{Liouvillea}
\end{equation}
Thus the Andreev level dynamics is described by a Liouville equation similar
to ordinary quantum mechanical two-level systems.

%%%%%%%%%%%%%%%%%%%%%%%%%%%%%%%%%%%%%%%%
\subsection{Andreev level current}

We conclude this section with derivation of an effective current operator for
the Andreev levels. A common quantum mechanical expression for the current
applied to the fields $\psi_{L,R}$ at $x = \mp 0$ gives,
\begin{equation}
I_{L,R}(t) = {e \over 2 m} \left[ \bar{\psi}_{L,R}(0,t) \left( - i \hbar
\partial_x \right) \psi_{L,R}(0,t) + h.\,c. \right] \,,
%%%\label{}
\end{equation}
or using Eq. (\ref{eom1+2a}),
\begin{equation}
I(t) = I_{L,R}(t) = (i e/ \hbar) \sqrt{D} \, \bar{\eta}(t) \, \sigma_z e^{- i
\sigma_z \phi(t) / 4} \psi_L(0,t) \,  + h.\,c. 
\label{I_LR}
\end{equation}
[the current is continuous, $I_L(t) = I_R(t)$, by virtue of Eq.
(\ref{eom3})]. The same equation can be obtained by varying the electronic
part of the action, $S_e$, with respect to the phase difference,
\begin{equation}
I(t) = {- 2 e \over \hbar} \frac{\delta S_e}{\delta \phi(t)}  \,. 
\end{equation}
Using this relation the statistically averaged current,
\begin{equation}
\langle I(t) \rangle = \int {\cal D}^2 \eta {\cal D}^2 \psi_{L,R} \;
 I(t)e^{i S_e/\hbar}\, ,
\end{equation}
can be written as
\begin{equation}
\langle I(t) \rangle = 2 e i \frac{\delta}{\delta \phi(t)} \int {\cal D}^2
\eta {\cal D}^2 \psi_{L,R} \; e^{i S_e/\hbar} \,, 
\label{Istatistical}
\end{equation}
or, after tracing out the fields, $\psi_{L,R}$,
\begin{equation}
\langle I(t) \rangle = 2 e i \frac{\delta}{\delta \phi(t)} \int {\cal D}^2
\eta  \; e^{i S_\eta^0/\hbar} \,. 
\label{Ieta}
\end{equation}
This leads to the following equation for the current in terms of the
Andreev variable,
\begin{equation}
I(t) = {- 2 e \over \hbar} \frac{\delta S^0_\eta }{ \delta \phi(t)} =
\bar\eta(t)\,I \,\eta(t)\,, 
\end{equation}
in which the kernel $I$ corresponds to the effective single particle current
operator of Andreev levels,
\begin{equation}
I = \frac{2e}{\hbar} \frac{d h_a}{d \phi} = - \frac{e}{\hbar} {\cal I}(\phi)
\, e^{-i \sigma_x \sqrt{R} \phi /2} \, \sigma_z \,,\;\;\; 
{\cal I}(\phi) = \Delta D \sin \frac{\phi}{2}. 
\label{calI}
\end{equation}
Apparently the current operator does not commute with the Andreev
level Hamiltonian, $[h_a\,,\,I]\neq 0$, which is the consequence of a
normal electron reflection at the QPC. Hence the Andreev level eigen
states consist of superpositions of the current eigen states, unless
$R=0$ (see Fig. 4). Correspondingly, the Andreev level current,
defined as an expectation value of the current operator over the
Andreev state,
\begin{equation}
I_a = \langle I \rangle_a = \pm {2e\over\hbar}{dE_a(\phi)\over d\phi}= 
\mp {eD\Delta^2\over 2\hbar E_a} \sin \phi,
\end{equation}
differs from the current eigenvalues, $\mp e{\cal I}/\hbar$ [$\langle ...
\rangle_a$ denotes averaging over the Andreev level eigenstate]. Thus the
Andreev level current undergoes strong quantum fluctuations. The spectral
density of current-current correlation function can be directly calculated by
using Eqs. (\ref{Ha}) and (\ref{calI}), [cf. Ref.\onlinecite{MRodero96}],
\begin{equation}
S(\omega) =  \langle I \rangle_a ^2 R \tan^2(\phi/2) \, 
\delta(\omega - 2E_a / \hbar).
%\label{}
\end{equation}

%%%%%%%%%%%%%%%%%%%%%%%%%%%%%%%%%%%%%%%%%%%%%%%%%%%%

\section{Averaging over phase fluctuations}

Equation (\ref{Liouvillea}) describes dynamics of the Andreev levels
for a given realization of the time dependent phase across the QPC.
However, the phase dynamics is strongly coupled to the Andreev
levels. Intrinsic dynamics of the phase is governed by the quantum
Hamiltonian of the loop oscillator [cf. Eq. (\ref{Losc})],
\begin{equation}
H_{osc} = {p^2\over 2 M} + {M \omega_p^2 \tilde{\phi}^2 \over 2}
\;,\;\; [\, p , \phi \, ] = - i \hbar \;,\;\; M = {\hbar^2\over 8E_C}, 
\label{Hosc}
\end{equation}
where $\omega_p = \sqrt{8E_LE_C /\hbar^2}$ is a plasma frequency of
the oscillator determined by the contact charging energy,
$E_C=e^2/2C$, and the loop inductive energy, $E_L =(\hbar
c/2e)^2(1/L)$. Thus the whole system is generally a multilevel one.
On the other hand, the phase dynamics and the coupling between the
loop oscillator and the Andreev levels is a vital part of the ALQ
operation: The Andreev levels cannot be manipulated if the phase
dynamics was frozen because any manipulation requires variation of
the current, and second, the read out of the Andreev levels can be
only performed via measuring the quantum state of the loop
oscillator. An obvious way to solve this problem and preserve the
qubit property of the coupled system is to "enslave" the loop
oscillator by choosing the oscillator level spacing $\hbar\omega_p$
much larger than the Andreev level spacing, $\hbar\omega_p \gg 2E_a$.
Then the Andreev state evolution will not excite the oscillator,
which will remain in the ground state and adiabatically follow the
evolution of the Andreev levels. This implies that the phase is a
fast variable which should be averaged out, leading to an effective
qubit Hamiltonian.

%%%%%%%%%%%%%%%%%%%%%%%%%%%%%%%%%%%%
\begin{figure}[h]
\centerline{\psfig{figure=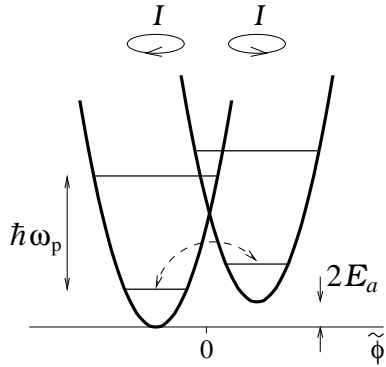,width=5cm}} \label{fig:average}
\vspace{.5cm} \caption { Potential energy diagram for two displaced
oscillator states corresponding to the different current states in the point
contact (shown as arrowed circles) . The plasma frequency, $\omega_p$, is
large compared to the Andreev level spacing; the oscillator remains in the
ground state during the qubit evolution (dashed arrow). }
\end{figure}
%%%%%%%%%%%%%%%%%%%%%%%%%%%%%%%%%%%%%%

To facilitate the averaging procedure we take advantage of the small
amplitude of the phase fluctuations, $\tilde{\phi} = \phi - \phi_e
\ll \phi_e$, which was already assumed when proceeding from Eq. (\ref
{Seta0_Fourier}) to Eq. (\ref {Seta0_slow}). This assumption is
justified by the large inductive energy, $E_L \gg E_J \approx
\Delta$, and it allows us to expand the Andreev level Hamiltonian
(\ref{Ha}) over small $\tilde\phi$; then proceeding to the current
eigen basis, $\eta \rightarrow e^{- i \sigma_x \sqrt{R} \phi_e/4}
\eta$, we get,
\begin{equation}
h_a = \Delta \,  \left( \cos \frac{\phi_e}{2} \, \sigma_z + \sqrt{R}
\sin \frac{\phi_e}{2} \, \sigma_y \right) - \frac{{\cal
I}(\phi_e)}{2} \, \tilde{\phi} \, \sigma_z \, = \, h^0_a + h_{int}\,.
\label{H:alq}
\end{equation}

Averaging over fast phase fluctuations can be done directly in Eq.
(\ref{U}) by performing explicit integration over phase. However,
there is a simpler way to get the same result. Equation
(\ref{Liouvillea}) holds for a fluctuating phase provided the
oscillator is not excited. It can be viewed as the diagonal with
respect to $\phi$ part of a more general equation for a full density
matrix, $\rho(\sigma\phi,\sigma'\phi')$, for the Andreev levels plus
oscillator system,
\begin{equation}\label{Lioville_full}
i\hbar \partial_t\rho = [h_a^0 + h_{int} + H_{osc}\,,\,\rho].
\end{equation}
The interaction between the Andreev levels and the oscillator given by the
second term in Eq. (\ref{H:alq}) displaces the oscillator steady state from
the origin by $\pm {\cal I}/2M\omega_p^2$ depending on the direction of the
current in the junction or, equivalently, the state of the Andreev levels
(see Fig. 5). We eliminate this term in (\ref{H:alq}) by applying the
transformation,
\begin{equation}
h_a \rightarrow e^{i A p } h_a \, e^{-i A p } \;,\;\; A = {{\cal
I}(\phi_e)\sigma_z \over 2M \hbar \omega_p^2} \,, 
\label{H:alq2}
\end{equation}
and then average the resulting Hamiltonian over the oscillator ground
state, taking into account the relation, ${\langle e^{i A \, p}
\rangle}_0 = \exp \left( - A^2 \, {\langle p^2 \rangle}_0 /2\right) $
[${\langle ...\rangle}_0$ indicates averaging over the oscillator
ground state]. As the result we get an effective Hamiltonian,
\begin{equation}
{\langle h_a \rangle}_0 = \Delta \left( \cos \frac{\phi_e}{2} \, \sigma_z
 + \sqrt{R^\ast} \sin \frac{\phi_e}{2} \,
\sigma_y \right) = h_q(\phi_e),
\label{Heff}
\end{equation}
where the bare contact reflectivity, $R$, is renormalized by the
phase fluctuations,
\begin{equation}
R^\ast = e^{-2 \lambda} R,\;\;\; \lambda = {{\cal I}^2(\phi_e) / 4 M
\hbar \omega_p^3}. 
\label{lambda}
\end{equation}
 This renormalization effect can be understood as the effect of
inertia of the loop oscillator, which hinders the current variations,
i.e. it works against the effect of the electronic back scattering at
the contact responsible for the hybridization of the current states
[see discussion in the end of the Section IIID]. The renormalization
effect becomes increasingly strong in the limit of a classical
oscillator with large ''mass''. Because of renormalization of the
contact reflectivity, the Andreev level spectrum is modified,
\begin{equation}
E_a (\phi_e) \rightarrow E_a^\ast (\phi_e) = \Delta \sqrt{ \cos^2(\phi_e/2) +
R^\ast \sin^2(\phi_e/2)} \,, 
\label{renormalization}
\end{equation}
and the qubit frequency reduces. This might be important for
practical applications, because it would allow one to reduce the
qubit frequency by choosing the circuit parameters rather than by
tuning the contact reflectivity.

Eq. (\ref{Heff}) gives effective Hamiltonian $h_q$ for the ALQ in the absence
of interaction with phonons. Similarly, averaged over phase fluctuation
density matrix $\langle \rho\rangle_0$ gives density matrix for the ALQ.
Keeping the same notation, $\rho$, for the qubit density matrix, we finally
arrive at the equation for the free evolution of the ALQ,
\begin{equation}
i\hbar \partial_t\rho = [h_q(\phi_e)\,,\,\rho]. \;\;\;
\label{ALQ_Liouville}
\end{equation}
This equation is sufficient to describe the ALQ manipulation (by driving the
biasing flux) and read out (by measuring the induced
current)\cite{Lantz02,Zazunov03}, and also the qubit-qubit interaction, which
is discussed in the next section.

% % % % % % % % % % % % % % % % % % % % % % % % % % % % % % % % % %

\section{Inductive qubit-qubit coupling}

Our treatment of the interaction of the Andreev levels with phase
fluctuations can be easily extended on the case of several
inductively coupled SQUIDs to describe direct qubit-qubit coupling.
Let us consider as an example two SQUIDs with different QPC
reflectivities, $R_{1}\neq R_2$, and identical circuit parameters,
$C$ and $L$, mutual inductance is $\cal M$. Then the inductance terms
in the Lagrangian (\ref{Losc}) written for the two qubits will take
the form,
\begin{equation}
{1\over 2}\left({\hbar c \over 2e}\right)^2\,
\tilde\phi^T \, \hat{L}^{-1} \tilde\phi
\;,\;\;\;
\phi = \left(
\begin{array}{c}
        \tilde\phi_1 \\
        \tilde\phi_2
\end{array}
\right)
\;,\;\;\;
\hat{L} = \left(
\begin{array}{cc}
        L & {\cal M} \\
        {\cal M} & L
\end{array}
\right) \,,
 %%\label{}
\end{equation}
where $\tilde\phi_{1,2}$ are fluctuating phases in the first and
second qubits. By introducing the normal modes for the
LC-oscillators, $\phi \rightarrow e^{-i \tau_y \pi/4} \phi$, we
present the two-qubit Hamiltonian on the form similar to Eqs.
(\ref{H:alq}), (\ref{H:alq2}),
\begin{equation}
H = \sum_{i=1,2} \left[\Delta \,  \left( \cos\frac{\phi_{e,i}}{2} \,
\sigma_{zi} + \sqrt{R_i} \sin \frac{\phi_{e,i}}{2} \,
  \sigma_{yi}\right)
 - \frac{\tilde\phi_i}{2\sqrt{2}} \,
\left( \,  (- 1)^{i+1} {\cal I}_1 \, \sigma_{z1} + {\cal I}_2 \,
 \sigma_{z2} \, \right) \, + \, H_{osc,i} \right]\,, 
\label{H12:qbinter}
\end{equation}
where $H_{osc,i}$ describes the normal oscillator with frequency $\omega_{pi}
= c/\sqrt{C \, (L \pm {\cal M})}$, and the index $i=1,2$ refers to the first
and second qubit. Then we apply the similar transformations as in the
previous section to Eq. (\ref{H12:qbinter}), namely, we eliminate the linear
in $\tilde\phi_{i}$ terms by means of a canonical transformation, $H
\rightarrow \exp \left( i \sum_{i} A_i p_i \right) H \exp \left( -i \sum_{i}
A_i p_i \right)$, with $A_i = \left( (-1)^{i+1} {\cal I}_1 \sigma_{z1} +
{\cal I}_2 \sigma_{z2} \right) / (2 \sqrt{2} M \hbar \omega_i^2) $, and then
average the transformed Hamiltonian over the ground state of the normal
oscillators. As the result, we obtain an effective two-qubit Hamiltonian
including direct qubit-qubit coupling,
\begin{equation}
H = h_{q1} + h_{q2} + (e/\hbar c)^2 \, {\cal M} \, {\cal I}_1 {\cal
I}_2 \sigma_{z1} \sigma_{z2}. 
\label{Hint}
\end{equation}
The renormalized contact reflectivities are now given by equation,
$R_i^\ast = R_i \exp[-\left( {\cal I}_{i}^2/4 \hbar M \right)
\left( \omega_1^{-3} + \omega_2^{-3} \right)]$.

It is worth mentioning that the two-qubit configuration may be also realized
with a {\em single} SQUID containing QPC with two conducting modes. In this
case, we have the two Andreev level Hamiltonians coupled to the same loop
oscillator,
\begin{equation}
H = \sum_{i=1,2} \left( h_{a,i}^0 - \frac{{\cal I}_i(\phi_e)}{2}
\tilde{\phi} \, \sigma_{zi} \, + \,H_{osc} \right)\,.
 %%\label{}
\end{equation}
Averaging over the phase fluctuations, we arrive at the same
interaction Hamiltonian as in Eq. (\ref{Hint}) but with the loop
inductance $L$ substituting for $-{\cal M}$. Thus we come to an
interesting conclusion that the effect of the phase fluctuations not
only reduces the bare contact reflectivity but also introduces
effective coupling of the Andreev levels of different conducting
modes in multimode QPC.

% % % % % % % % % % % % % % % % % % % % % % % % % % % % % % % % % % % % %
% % % % % % % % % % % % % % % % % % % % % % % % % % % % % % % % % % % % % %

\section{Andreev level - phonon interaction}
\subsection{Effective action}

In this Section we include electron-phonon interaction in the
consideration. We start with the derivation of an effective action
for the Andreev level-phonon interaction. To this end we repeat the
calculation of the previous section adding the Lagrangian $L_{e-ph}$,
Eq. (\ref{Le-ph}), to Eq. (\ref{intpsi}). By keeping in the electrode
Green functions only the first-order correction in small interaction
constant $\gamma$, we arrive at the following action,
\begin{equation}
S_\eta = - \int dt_1 dt_2 \, \bar{\eta}(t_1) \left[ D e^{-i \sigma_z
\phi(t_1)/4} G_L(t_1, t_2) \, e^{i \sigma_z \phi(t_2)/4} +  (1+\sqrt{R})^2 \,
e^{i \sigma_z \phi(t_1)/4} G_R(t_1, t_2) e^{-i \sigma_z \phi(t_2)/4} \right]
\eta(t_2) \,, 
\label{Seta}
\end{equation}
where the Green functions $G_{L,R}$ read,
\begin{equation}
G_\alpha(t_1, t_2) = \sum_{\sigma = \pm} \left[ \,
g_\alpha^{(\sigma)}(0,0;t_1 - t_2) + \gamma \int d {\bf r} \, |\psi_\perp|^2
\int dt \, g_\alpha^{(\sigma)}(0,x; t_1 - t) \, \nabla {\bf u}({\bf r},t) \,
\sigma_z  g_\alpha^{(\sigma)}(x,0; t-t_2) \, \right] \,. 
\label{G_LR}
\end{equation}
In this equation, the quantities $g_{L,R}^{(\sigma)}$ refer to the different
parts, $ g^{(\sigma)} (x<0, x'< 0;t)$ and $g^{(\sigma)} (x>0, x' > 0;t)$,
respectively, of the translation-invariant free electron Green function,
$g^{(\sigma)}(x-x',t)$, which obeys the equation
\begin{equation}
\left[ i \hbar \partial_t + \sigma \sigma_z v  i \hbar \partial_x - \Delta \,
\sigma_x \right] \, g^{(\sigma)} (x - x',t) = \delta(x - x') \, \delta(t) \,.
%%\label{}
\end{equation}
The Green functions $g_{\alpha} ^{(\sigma)}$ in Eq. (\ref{G_LR}) are
explicitly given by
\begin{equation}
g_{L,R}^{(\sigma)}(x,0;t) = - \Theta(\mp x) \,
\frac{e^{-|x| \zeta_t/ \hbar v}}{2 \hbar v \, \zeta_t} \, \left[ \, i \hbar
\partial_t + \Delta \, \sigma_x \mp \sigma \, i \, \zeta_t \, \sigma_z
\right] \, \delta(t) \,,
 %%\label{}
\end{equation}
where $\zeta_t$ is given by the Fourier component, 
$\zeta_\omega = \sqrt{\Delta^2 - (\hbar \omega)^2}$.

Proceeding to the adiabatic approximation discussed in the previous section
($\zeta_\omega \rightarrow \zeta_e$), and performing the transformation
(\ref{transformation}), we arrive at the following effective action for the
Andreev level-phonon interaction,
\begin{equation}
S_{\eta-ph} = - \gamma \int dt \int d {\bf r} \, |\psi_\perp|^2 \, n(x,t) \,
\nabla{\bf u}({\bf r},t) \,, 
\label{Seta-ph}
\end{equation}
where
\begin{equation}
n(x,t) = \frac{e^{-2 |x| \zeta_e/ \hbar v}}{4 \hbar v \, \zeta_e} \, (1 +
{\rm sgn}x \, \sqrt{R}) \, \Lambda(x,t) \,, 
\label{nnn}
\end{equation}
and
\begin{equation}
\Lambda(x,t) = \hbar^2 \left(\partial_t \,\bar{\eta} \right) \sigma_x
\left(\partial_t \eta \right) + \hbar \Delta \left( \partial_t \bar{\eta} \,
 e^{- i \sigma_x (\sqrt{R} - {\rm sgn} x ) \phi_e/2} \sigma_y \eta +
h.c. \, \right) - E_a^2 \, \bar{\eta} \,\sigma_x \eta .
%%\label{}
\end{equation}
Taking into account the zero-order dynamic equation with respect to
$\gamma$, Eq. (\ref{equation_eta}), and putting $ \phi = \phi_e$,
\begin{equation}
i \hbar \partial_t \eta = \Delta \, e^{-i \sigma_x  \sqrt{R} \, \phi_e /2}
\left( \cos \frac{\phi_e}{2} \, \sigma_z + \sqrt{R} \, \sin \frac{\phi_e}{2}
\, \sigma_y \right) \eta \,,
 %%\label{}
\end{equation}
we obtain for the quantity $n(x,t)$, Eq. (\ref{nnn}), the following
expression,
\begin{equation}
n(x,t) = (\sqrt{R}/2) \, \kappa \, {\rm sgn} x \, e^{- 2\kappa |x|} \,
\bar{\eta} \, \sigma_x \eta \,,
%%\label{}
\end{equation}
where $\kappa =  \zeta_e / \hbar v$. Finally, integrating over ${\bf r}$ in
the action (\ref{Seta-ph}), we present the effective interaction on the form,
\begin{equation}
S_{\eta-ph} = - \int dt \, \sum_{\bf q} \, \gamma_{\bf q} X_{\bf q}
\bar{\eta} \, \sigma_x \eta \,, 
\label{Seta-ph:final}
\end{equation}
where $\gamma_{\bf q}$ is the effective constant of the Andreev level-phonon
coupling,
\begin{equation}
\gamma_{\bf q} = \gamma \kappa \, \sqrt{{ \hbar \Omega_q R \over 2
\rho V s^2}} \int_0^{+ \infty} dx \, F ({\bf q}_\perp, x) \, e^{-2
\kappa x} \sin(q_x x) \,,\;\;\; F ({\bf q}_\perp, x) = \int d {\bf
r}_\perp \, |\psi_\perp(\bf r)|^2 \, e^{i {\bf q}_\perp {\bf
r}_\perp} \,. 
\label{gammaq}
\end{equation}
We notice that the Andreev level-phonon interaction in Eq.
(\ref{Seta-ph:final}) has purely transverse origin, i.e. while
inducing interlevel transitions and hence the relaxation it does not
produce any additional dephasing to the one associated with the
relaxation.

It is important to mention that the effective coupling constant is
proportional to $\sqrt R$, and it turns to zero in the case of perfectly
transparent constriction. This results from the already mentioned fact that
the relevant phonons have small wave vectors and are not able to provide
large momentum transfer ($\sim 2\hbar k$) during scattering with electrons.

The effective action in Eqs. (\ref{Seta-ph:final}) and (\ref{gammaq}) was
derived for a given realization of the time dependent phase. To take into
account the effect of phase fluctuation, we have to apply the transformation,
Eq. (\ref{H:alq2}), to the action (\ref{Seta-ph:final}); the Lagrange form of
the transformation reads,
\begin{equation}
\eta \rightarrow \exp\left(i \,{{\cal I}(\phi_e)\sigma_z\over 2\hbar} \int^t
d\tau \tilde\phi(\tau) \right)\eta \,. 
\label{H:alq2action}
\end{equation}
Then the integration over the phase adds the factor $e^{-\lambda }$ to the
action, cf. Eq. (\ref{lambda}), which simply implies a renormalization of the
coupling constant,
\begin{equation}
\gamma_{\bf q} \rightarrow \gamma_{\bf q}^\ast = e^{-\lambda}
\gamma_{\bf q} \,. 
\label{gamma_star}
\end{equation}
This result can be expected: since the Andreev level-phonon coupling
is transverse in the current basis and depends on the contact
reflectivity, the renormalized reflectivity $R^\ast$ rather than the
bare one has to enter the coupling constant, $\gamma_q^\ast =
\gamma_q (R^\ast)$; this is consistent with Eqs. (\ref{gamma_star}),
(\ref{gammaq}).

%%%%%%%%%%%%%%%%%%%%%%%%%%%%%%%%%%%%%%%%%

\subsection{Kinetic equation}

Qubit decoherence is usually described with collision terms in the Liouville
equation for the qubit density matrix, which take into account interaction
with an environment. Our goal now will be to derive the phonon-induced
collision terms in the equation (\ref{ALQ_Liouville}) for ALQ density matrix,
and to evaluate the decoherence of the ALQ.

While the description of free qubit evolution was possible in the terms of
the single-particle density matrix, evaluation of the collision terms goes
beyond the single-particle approximation and requires the knowledge of {\em
electronic} two-particle correlation functions. This is because the Andreev
levels are not a rigorously isolated system but rather belong to a large
fermionic system of the superconducting electrons in the contact electrodes.
Thus to derive the collision terms, we apply the many-body Keldysh-Green
function technique\cite{Landau} combined with the path integral approach. The
method described below automatically takes into account the many-body
effects, namely the Pauli exclusion principle, leading to a non-linear form
of the collision terms and eventually to suppression of the decoherence.

The starting point of the derivation is Eqs. (\ref{U}), (\ref{Ue})  for the
propagator, in which the integration over the fast fermionic fields,
$\psi_{L,R}$, and phase, $\phi$, has been performed while integration
over the phonons and Andreev states remains,
\begin{equation}
{\cal U} = \int {\cal D} \{ X_{\bf q} \} \,
{\cal D}^2 \eta \, e^{i S / \hbar}
\,.
\label{Unew}
\end{equation}
The time evolution in the action follows now along the Keldysh
contour\cite{Keldysh} $C_K$, $S = \int_{C_K} dt \, L$, which goes from $-
\infty$ to $+ \infty$, and then backwards.\cite{Zaikin90,Kamenev99,Golub01}
The interaction is supposed to be switched on and off adiabatically at the
remote past, $t = - \infty$,  and the phonon bath is supposed to be in the
thermal equilibrium.
The Lagrangian $L$ has the form,
\begin{equation}
L = \bar{\eta}
\left( i \hbar \partial_t - E_a^\ast \sigma_z \right) \eta -
\sum_{\bf q}
\gamma_{\bf q}^{*} \, X_{\bf q} \, \bar{\eta} \, \sigma_x \eta  + L_{ph}
\,,
\label{L}
\end{equation}
where $ L_{ph}$ is given by Eq. (\ref{Lph}). To reduce the time integration
along the Keldysh contour to an ordinary time integral, we distinguish
forward and backward branches of the contour by labeling them with index
$s=1,2$, and introduce the two-component fields $\eta^s$ and $X_{\bf q}^s$.
Since the action is local in time, it can be rewritten as $S =
\int_{-\infty}^{+\infty} dt \, \left( L^1 - L^2 \right)$, where $L^s = L[\,
\bar{\eta}^s, \eta^s, X_{\bf q}^s]$.

The first step of the derivation is to integrate out the phonon
fields, which will give rise to an effective self-interaction
for the field $\eta$,
\begin{equation}
S_{int}[\bar{\eta}, \eta] = - {1 \over 2}
\int dt dt' \,
\left[ \bar{\eta}(t) \sigma_x \eta(t) \right]^{s}
\left[ \tau_z \check{D}(t - t') \tau_z \,
\right]^{s s'}
\left[ \bar{\eta}(t') \sigma_x \eta(t') \right]^{s'}
\,,
\label{Sint}
\end{equation}
with the kernel $\check{D}(t - t')$ given by
\begin{equation}
\check{D}(t - t') =
\sum_{\bf q} \, |\gamma_{\bf q}^{*}|^2 \,
\check{D}_{\bf q}(t - t') \,, \;\;
D_{\bf q}^{s s'}(t - t') = -(i/\hbar)
\langle T_C \left( X^{s}_{\bf q}(t) \,
X^{s'}_{\bf q}(t') \right) \rangle \,,
\label{D}
\end{equation}
where $\check{D}_{\bf q}(t - t')$ is the equilibrium Keldysh Green
function of the phonons represented with the $2 \times 2$
matrix in the Keldysh space.
In these equations, $T_C$ is the time-ordering operator on the
Keldysh contour, and $\tau_z$ is the Pauli matrix operating
in the Keldysh space; summation over repeated indices is implied.

The next step is to take advantage of the weak electron-phonon interaction, and
to decouple the four-fermionic interaction term in Eq. (\ref{Sint}) by
introducing the Hubbard-Stratonovich field ${\cal G}^{s s'}_{\alpha
\beta}(t,t')$, which is a matrix in the Keldysh-Nambu-time space. Before
doing this, it is convenient to explicitly extract the small parameter
$\lambda_{ph}$, which determines the electron-phonon coupling strength, from
the kernel $\check{D}$ in Eq. (\ref{Sint}) by redefining the kernel,
$\check{D} \rightarrow \lambda_{ph} \check{D}$ ($\lambda_{ph} \sim (v/s)
(\omega / \omega_D)^2 \ll 1$, $\omega_D$ is the Debye frequency). As the
result, we get,
\begin{equation}
{\cal U} = \int {\cal D} \, \check{{\cal G}} \, \, {\cal D}^2 \eta \,
\exp \left[ (i / \hbar) \int dt dt' \, \bar{\eta}^{s}(t)
\left[ {\cal L}(t,t') \tau_z -
\tau_z \check{\Sigma}(t,t' ) \tau_z \right]^{s s'}
\eta^{s'}(t') + i W[\, \check{{\cal G}}\, ] \, \right] \,,
\label{HS}
\end{equation}
where
${\cal L}(t,t') = \left( i \hbar \partial_{t} - E_a^\ast \sigma_z
\right) \delta (t - t')$ is diagonal in the Keldysh space, and
\begin{equation}
W[\, \check{{\cal G}} \, ] =  {\hbar \over 2 \lambda_{ph}}
\int dt dt' \,
{\cal G}^{s s'}_{\alpha \beta}(t,t') \, \left[ \tau_z \check{D}(t - t')
\tau_z \right]^{s s'} \left[ \sigma_x \, \check{{\cal G}}(t', t) \, \sigma_x
\right]^{s' s}_{\beta \alpha} \,,
%%\label{}
\end{equation}
\begin{equation}
\Sigma^{s s'}_{\alpha \beta}(t, t') = i \hbar D^{s s'}(t -t')
\left[ \sigma_x \, \check{{\cal G}}(t, t') \, \sigma_x
\right]^{s s'}_{\alpha \beta} \,.
\label{self-energy}
\end{equation}
Equation (\ref{HS}) describes dynamics of the field $\eta$ interacting with
the effective field $\check{{\cal G}}$. In terms of the Keldysh-Green function
for the field $\eta$,
\begin{equation}
G^{s s'}_{\alpha \beta}(t,t') = -{(i/\hbar) \langle T_C \left(
\eta^{s}_\alpha(t) \, \eta^{\dagger s'}_\beta(t') \right) \rangle} \,,
\label{Geta}
\end{equation}
this evolution is described by the Dyson equation,
\begin{equation}
\left( {\cal L} \tau_z - \tau_z \check{\Sigma} [{\check{\cal G}}] \tau_z
\right) \check{G} = \check{1} \,,
\label{DysonG}
\end{equation}
where the self-energy $\check{\Sigma}$ depends on the effective field
$\check{{\cal G}}$. A closed equation for $\check{{\cal G}}$ can be derived
by integrating out the field $\eta$ in Eq. (\ref{HS}). This procedure leads
to the equation,
\begin{equation}
{\cal U} = \int {\cal D} \check{{\cal G}} \, e^{i S[\,\check{{\cal G}}\,] /
\hbar} \;,\;\;\;
S[\check{{\cal G}}]/\hbar = - i {\rm Tr} \ln \left( {\cal L}
\, \tau_z - \tau_z \check{\Sigma} [\check{\cal G}] \tau_z \right) +
W[\,\check{{\cal G}}\,] \,, 
\label{UG}
\end{equation}
where ${\rm Tr}$ denotes both the matrix trace in the Keldysh-Nambu space and
the integration over the time variables. Noticing that the action is large,
$S[\,\check{{\cal G}}\,] \sim \lambda^{-1}_{ph}$,
we evaluate the integral
in Eq. (\ref{UG}) within the saddle-point approximation
(cf. Ref. \onlinecite{Kamenev99}).
The corresponding saddle-point equation is derived
by varying the action with respect to $\check{{\cal G}}$, which yields,
\begin{equation}
\left( {\cal L} \tau_z - \tau_z \check{\Sigma} [\check{{\cal G}}] \tau_z
\right)
\check{{\cal G}} = \lambda_{ph} \check{1}.
\label{DysonCalG}
\end{equation}
Comparing Eqs. (\ref{DysonCalG}) and (\ref{DysonG}), we obtain the relation,
$\check{\cal G} = \lambda_{ph} \check{G}$. Written in the terms of
$\check{G}$,
\begin{equation}
\left( {\cal L} \tau_z - \tau_z
\check{\Sigma} [\lambda_{ph} \check{G}] \, \tau_z
\right) \check{G} = \check{1} \,,
\label{Dyson4G}
\end{equation}
the saddle-point equation is the Dyson equation for the qubit Keldysh-Green
function, Eq. (\ref{Geta}), in which the self-energy contains only an
undressed vertex part and a free phonon Green function. Including the
parameter $\lambda_{ph}$ back into the kernel, $\check{D}$, $\lambda_{ph}
\check{D} \rightarrow \check{D}$, we arrive at the expression for the
self-energy, Eq. (\ref{self-energy}), with $\check{\cal G}$ being replaced by
$\check{G}$, while $\check{D}$ is given by Eq. (\ref{D}).

To proceed with the derivation of kinetic equation, it is convenient to
introduce a triangular form for the Keldysh-Green function by performing
transformation in the Keldysh space,
\begin{equation}
\check {G} \rightarrow \check{L} \tau_z \check{G} \check{L}^{-1} =
\left(
\begin{array}{cc}
G^R & G^K \\
0 & G^A
\end{array}
\right)
\;,\;\;\;
\check{L} = {1 \over \sqrt{2}}
\left(
\begin{array}{rr}
1 & -1 \\
1 & 1
\end{array}
\right) \,,
\label{G_}
\end{equation}
where $G^{R(A)} = G^{11} - G^{12(21)}$ is the retarded
(advanced) Green function, and $G^K = G^{11} + G^{22} =
G^{12} + G^{21}$ is the Keldysh component.
Similar relations also hold for the self-energy.
Then Eq. (\ref{Dyson4G}) takes the form,
\begin{equation}
\check{{\cal L}} \, \check {G} =
\check{1} + \check{\Sigma} \, \check {G} \,.
\label{Deq}
\end{equation}
Kinetic equation is obtained by considering the difference between
Eq. (\ref{Deq}) and its Hermitian conjugate for the Keldysh
component,\cite{Landau}
\begin{equation}
i \hbar \left( \partial_t + \partial_{t'} \right) G^K(t,t') - E_a^\ast
\left[ \, \sigma_z ,  G^K(t,t') \, \right] =
\left( \Sigma^R G^K - G^K \Sigma^A +
\Sigma^K G^A - G^R \Sigma^K \right) (t,t') \,.
\label{GK_eq_time}
\end{equation}
%
%where $\otimes$ denotes convolution in time space.

The right hand side of Eq. (\ref{GK_eq_time}) describes the qubit
decoherence as well as dynamic corrections due to the
phonons.\cite{RammerSmith} In the absence of the coupling to phonons,
the solution of Eq. (\ref{GK_eq_time}) has the form,
\begin{equation}
G^K_0(t,t') = (-i / \hbar) \, e^{-i  E_a^\ast \sigma_z t /\hbar} \, F \,
e^{i  E_a^\ast \sigma_z t' /\hbar} \,,
%%\label{}
\end{equation}
where $F$ is a  time-independent matrix determined by the initial state of
the qubit. When a weak interaction with the phonons is switched on, an
asymptotical solution to the equation (\ref{GK_eq_time}) can be written on
the form,
\begin{equation}
G^K(t,t') =
(-i / \hbar) e^{-i  E_a^\ast \sigma_z t /\hbar}
\left[  \, F((t + t')/2) + \tilde{F}(t,t') \, \right]
e^{i  E_a^\ast \sigma_z t' /\hbar} \,,
%%\label{}
\end{equation}
where the matrix $F$ is a slowly evolving on the time scale, $\hbar/ E_a^\ast$,
function of the global time $(t+t')/2$,  and $\tilde{F}$ is a rapidly
oscillating small-amplitude correction (see Appendix B). Equations for the
matrix elements of $F$ are derived in Appendix B, Eqs. (\ref{F1ap}),
(\ref{F12ap}), and for the diagonal matrix elements they read,
\begin{equation}
\partial_t F_1 = - \partial_t F_2 =
- {\nu \over 2} \left[ \,
(2N+1)\left( F_1 - F_2 \right) +  F_1 F_2  - 1
\, \right]
\,,
\label{F1}
\end{equation}
while for the off-diagonal matrix element, $F_{12}=F_{21}^\ast$,
the equation has the form,
\begin{equation}
\partial_t F_{12} = - \left[ \, {\nu \over 2} \,
\left( 2 N + 1 - F_z \right) +
2 i (\delta + \delta_0 F_z) \, \right] F_{12}
\;,\;\;\;
F_z(t) = \left( F_1 - F_2 \right)/2
\,,
\label{F12}
\end{equation}
where $\nu$ is the phonon-induced transition rate between the qubit levels,
\begin{equation}
\nu = ( 2 \pi / \hbar )
\int \frac{d^3 q}{(2 \pi)^3} \, V \, |\gamma_{\bf q}^{*}|^2 \,
\delta( 2 E_a^\ast - \hbar \Omega_q)
\,.
\label{nu0}
\end{equation}
$N = \left( e^{2\beta E_a^\ast} - 1 \right)^{-1}$ is the phonon distribution
function at the frequency equal to the qubit level spacing, $\beta = 1/kT $,
and $\delta$ and $\delta_0$ are small dynamic corrections defined in Eq.
(\ref{deltas}).

For equal times, $t = t'$, $G^K(t,t)$ is related to the qubit density
matrix, Eq. (\ref{rhoa}), as follows: $G^K_{\sigma \sigma'}(t,t) = (-i/\hbar)
\left[ \, 2 \rho_{\sigma \sigma'}(t) - \delta_{\sigma \sigma'} \right]$, and
therefore Eqs. (\ref{F1}), (\ref{F12}) in fact give kinetic equation for the
qubit density matrix in the interaction picture,
$\tilde\rho = e^{i \sigma_z E_a^\ast t/\hbar}\rho \, 
e^{-i \sigma_z E_a^\ast t/\hbar}$, $F_{\sigma} = 2 \rho_{\sigma\sigma} -1$,
$F_{12} = 2\tilde{\rho}_{12}$. It is instructive to write Eq. (\ref{F1}) 
in terms of the qubit occupation numbers $n_\sigma = 1- \rho_{\sigma \sigma}$,
\begin{equation}
\partial_t n_1 = - \partial_t n_2 =
- \nu \left[ \,
(N+1) \, n_1 (1-n_2) \, - \, N n_2 \, (1 - n_1)\, \right] \,.
\label{n1}
\end{equation}
The right hand side of this equation has the standard form of the
electron-phonon collision term, yielding the Fermi distribution for the
equilibrium occupation numbers,
\begin{equation}
n_{1,2}^0 = {1\over e^{\pm\beta E_a^\ast} + 1}.
\label{neq}
\end{equation}
This conclusion is consistent with the well known fact that the Fermi
distribution of the Andreev levels gives correct magnitude for the
equilibrium Josephson current\cite{Furusaki90,Beenakker91}. Furthermore, it
follows from Eqs. (\ref{n1}) and (\ref{neq}) that ${\rm Tr}\,\rho (t) =1$.
Then equations for the two independent components of the density matrix,
$\rho_z = (\rho_{11} - \rho_{22})/2$, and $\tilde{\rho}_{12}$, 
omitting the dynamic corrections, read,
\begin{equation}
\partial_t \rho_z =
- \nu \left[ \, \left( 2 N + 1 \right) \rho_z - \rho_z^2 - 1/4 \, \right]
\,, 
\label{rhoz_eq}
\end{equation}
\begin{equation}
\partial_t \tilde{\rho}_{12} = - \nu \left( N + 1/2 - \rho_z \right) 
\tilde{\rho}_{12} \,.
\label{rho12_eq}
\end{equation}
These nonlinear equations are drastically different from the linear
Bloch-Redfield equation describing decoherence of the macroscopic
superconducting qubits,\cite{Shnirman} and they have qualitatively different
solutions, as illustrated in Fig. 6.
Exact solutions for Eqs. (\ref{rhoz_eq}), (\ref{rho12_eq}) read,
\begin{equation}
\delta\rho_z(t) = \frac{\delta\rho_z(0) \, e^{- \Gamma t}}
{1 + \delta\rho_z(0) \sinh (\beta E_a^\ast) \left( 1 - e^{- \Gamma t} \right)}
\,,\;\;\;
\Gamma = {\nu \over \sinh (\beta E_a^\ast)} \, ,
\label{rhoz}
\end{equation}
\begin{equation}
\tilde{\rho}_{12}(t) =  
{\rho_{12}(0) \, e^{- \Gamma t/2} \over 1 + \delta\rho_z(0)
\sinh (\beta E_a^\ast) \left( 1 - e^{- \Gamma t} \right)} \,, 
\label{rho12}
\end{equation}
where $\delta\rho_z(t) = \rho_z^0 - \rho_z(t)$ is the deviation from the
equilibrium, $\rho_z^0 = (1/2) \tanh (\beta E_a^\ast / 2)$. The evolutions of
the diagonal (relaxation) and off-diagonal (dephasing) parts of the density
matrix are qualitatively similar. One may distinguish the linear regime,
$\delta\rho_z(0) \sinh (\beta E_a^\ast)\ll 1$, when the decoherence is
determined by the exponential law,
\begin{equation}
\delta\rho_z(t) = \delta\rho_z(0) \, e^{- \Gamma t}, \;\;
\tilde{\rho}_{12}(t) = \rho_{12}(0) \, e^{- \Gamma t /2}.
\label{exp}
\end{equation}
%\noindent
However, the decoherence rate, $\Gamma$, becomes exponentially
small at temperature smaller than the qubit level spacing, $\beta E_a^\ast
\gg 1$. At this temperature, the most interesting is the opposite, nonlinear
regime, $\delta\rho_z(0) \sinh (\beta E_a^\ast)\gg 1$. In this case, there is
a wide time interval, $t \ll \sinh (\beta E_a^\ast)/\nu$, where both the
relaxation and dephasing follow the power law (see Fig. 6),
\begin{equation}
\delta\rho_z(t) = {1\over\nu t}, \;\;\;
\tilde{\rho}_{12}(t)= {\rho_{12}(0)\over\delta\rho_z(0)}{1\over\nu t},
\label{nexp}
\end{equation}
and only at very large times, $t \gg \sinh (\beta E_a^\ast)/\nu$, the
decoherence undergoes a crossover to an exponential regime similar to Eq.
(\ref{exp}). We note that the exponentially small relaxation rate in the
linear regime  is well known for the quasiparticle recombination in bulk
superconductors.\cite{Kaplan}
%
%%%%%%%%%%%%%%%%%%%%%%%%%%
\begin{figure}[t]
\centerline{\psfig{figure=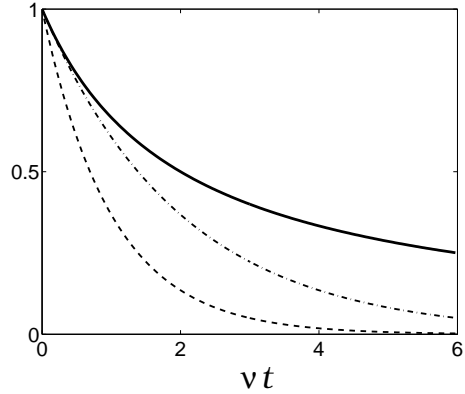,width=6.1cm}} \vspace{.5cm} \caption{
Decay with time of a "cat" state ($\delta\rho_z(0)=\rho_{12}(0) = 1/2$). Bold
line indicates evolution of both (normalized) density matrix elements for ALQ
for $1/\beta = 0.2 E_a^{\ast}$; for comparison, exponential relaxation and
dephasing of a macroscopic superconducting qubit is illustrated with the
dashed and dashed-dotted lines, respectively. }
\end{figure}
%%%%%%%%%%%%%%%%%%%%%%%%%

% % % % % % % % % % % % % % % % % % % % % % % % % % % % % % % % % % % % % %

\subsection{Evaluation of transition rate}

We conclude our study with the evaluation of the phonon-induced transition
rate, $\nu$ in Eq. (\ref{nu0}). To evaluate the transition rate, one needs
to specify geometry of the junction in more detail. Let us suppose that our
adiabatic constriction, Eq. ({\ref{Psi}), is formed by a hard-wall potential
and has an axial symmetry. Under these assumptions, the Fourier component of
transverse wave function in Eq. (\ref{gammaq}) has the form,
\begin{equation}
F({\bf q}_\perp,x) = 2 \,
{J_1(r_\perp(x) q_\perp) \over r_\perp(x) q_\perp},
\label{transvF}
\end{equation}
where $r_\perp(x)$ is the radius of the constriction cross section. The
magnitude of the relaxation rate essentially depends on the parameter
$r_\perp(0)Q$, where  $r_\perp(0)$ is the radius in the neck of the
constriction, and $Q = 2 E_a^\ast / \hbar s$ is the wave vector of phonons
responsible for the interlevel transitions; for atomic-size constrictions,
this parameter is small, $r_\perp(0)Q \ll 1$. Let us assume that the qubit
level spacing is not too small, $E^\ast_a/\Delta \,\gg\, s/v$, then the
phonon wave vector $Q$ is large compared to the inverse penetration length of
the Andreev level wave function, $Q \gg \kappa$.

Let us for a moment assume that the Andreev level wave function does not
spread out in the electrodes, but remains confined in the transverse
direction, $r_\perp(x) = const = r_\perp(0)$ (see Fig. 7); then the
Fourier component in Eq. (\ref{transvF}) is close to unity, and the
interaction region in Eq. (\ref{gammaq}) is limited by the penetration length
of the Andreev state, $x\sim 1/\kappa$, restricting relevant phonon
longitudinal wave vectors to small values, $q_x \sim \kappa \ll Q$. The
transition rate in this case reads,
\begin{equation}
\nu_0 = \frac{\gamma^2 R^\ast}{16 \hbar \rho s^2} \, \kappa Q^2 \sim
R^\ast (\zeta_e / E_a^\ast) \, \tau_{ph}^{-1}(E_a^\ast) \,,
\label{rateIF}
\end{equation}
where $\tau_{ph}^{-1}(E_a^\ast) \, \sim \, E_a^{*3}/\hbar \Theta^2_D$
is a bulk electron-phonon relaxation rate at the Andreev level energy
($\Theta_D$ is the Debye temperature).
This result has been derived in Ref. \onlinecite{IvanovFeig:ph}
(although neglecting the renormalization effect),
and it can be qualitatively applied to long constrictions,
whose length exceeds the coherence length. For short constrictions
considered here, $L \ll 1/\kappa$, the effect of spreading out of the Andreev
level wave function is essential, and  the approximation, $r_\perp = const$,
is not appropriate.

%%%%%%%%%%%%%%%%%%%%%%%%%%%%%%%%%%%%
\begin{figure}[h]
\centerline{\psfig{figure=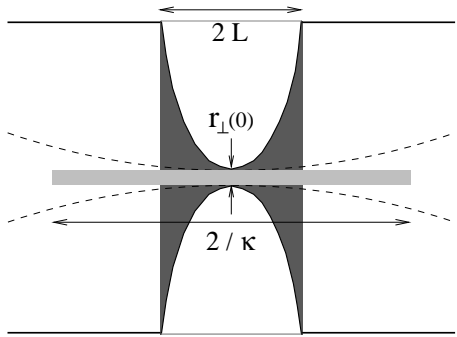,width=6cm}}
\vspace{.5cm}
\caption
{
Interaction region of the Andreev levels with phonons
in short QPC (dark shadow) and in long QPC (light shadow);
in long QPC, increase of the constriction radius (dashed line) can be neglected.
}
\end{figure}
%%%%%%%%%%%%%%%%%%%%%%%%%%

Let us adopt the following model for the constriction shape,
$ r_\perp(x) = r_\perp (0)\left( 1-|x|/L \right)^{-a} $, $a>1$.
It is easy to see then that the integral in
Eq. (\ref{gammaq}) will have a cut off at $|x| = \,L\,\ll \,1/\kappa$.
Function $J_1(z^a)/z^a$ in Eq. (\ref{transvF}) can with good accuracy be
approximated with the function, $(1/2) \, \theta(z_0 - z)$,
where $z_0 \approx 1$, and the integral in Eq. (\ref{gammaq})
is easily evaluated, giving,
\begin{equation}
\nu = \kappa L \nu_0 \,,
\label{rateK}
\end{equation}
i.e., the transition rate in short constrictions significantly reduces.
This is the effect due to small spatial region available
for the Andreev level-phonon interaction in short QPC.

%%%%%%%%%%%%%%%%%%%%%%%%%%%%%%%%%%%%%%%
\section{Conclusion}

Let us summarize the outlined theory for the Andreev level qubit (ALQ). ALQ
belongs to the family of superconducting flux qubits, but it differs from the
macroscopic flux qubits\cite{Delft2000,Friedman00,Delft2003,Vion02} in
several important respects. First, the quantum hybridization of the flux (and
persistent current) states in the ALQ loop is produced by electronic back
scattering in the quantum point contact (QPC) rather than by charge
fluctuations on the junction capacitors in the case of the macroscopic
qubits. Thus, in principle, neither small junction capacitance nor large loop
inductance is critical for the ALQ operation. Second, the ALQ is based on a
QPC with large, almost full, transparency, in contrast to classical tunnel
junctions employed in the macroscopic superconducting qubits. Large contact
transparency is required for placing Andreev levels deep within the
superconducting energy gap to achieve good decoupling from the continuum
electronic states. To guarantee good separation of the qubit levels from the
continuum, the amplitude of the phase fluctuations around the biasing point,
$\phi\approx\pi$, must be restricted to the small values, which implies small
inductance of the qubit loop.

In the tunnel junctions of the macroscopic qubits, Andreev levels are
fast variables whose effect, after averaging, reduces to the
Josephson potential energy added to the Hamiltonian of the loop
oscillator. In the transparent junction of the ALQ, Andreev levels
are slow variables which cannot be averaged out, and the full
description includes the Andreev two-level Hamiltonian strongly
coupled to the quantum loop oscillator. Derivation of the effective
two-level Hamiltonian goes beyond the tunnel model approximation,
which is done by incorporating exact boundary condition into the
action of the contact.

The qubit read out is achieved via measuring fluctuating persistent
current or induced flux in the qubit loop. To simplify the read out,
the loop plasma frequency is chosen to be large compared to the qubit
frequency so that the loop oscillator is "enslaved" by the Andreev
levels. Other regimes, e.g. resonance between the Andreev levels and
loop oscillator, could be also considered, however they remained
outside the scope of the paper. Typical circuit parameters for the
ALQ could be chosen as follows: $E_J\approx\Delta$ for the single
open mode, $R \leq 0.01$ with $I_c\sim$ 400nA for Nb; $L\sim$0.1nH,
and $C\sim $ 0.1pF, giving $\omega_p \sim 10^{11}$sec$^{-1}$, and
$E_L/\hbar\sim 10^{13}$sec$^{-1}\sim 10\Delta_{Nb}/\hbar$, while
$E_a^\ast/\hbar \sim 10^{10}$sec$^{-1}$.

Full description of the ALQ dynamics and qubit-qubit coupling,
including qubit manipulation and read out is given by the
single-particle density matrix.

Even if a good decoupling of the qubit states from the continuum electronic
states is provided, there are still soft microscopic modes in the junction,
which could couple to the Andreev levels. These modes present potential
source of "intrinsic" decoherence, in addition to the commonly considered
external decoherence, e.g., due to fluctuating biasing and read out circuits.
We considered such an intrinsic decoherence of the ALQ related to acoustic
phonons. It turns out that the collision terms in the kinetic equation are
nonlinear, in contrast to the linear master equation for the macroscopic
superconducting qubits.\cite{Shnirman} This reflects the fermionic nature of
the Andreev states and leads to considerable enhancement of the decoherence
time at small temperature. One can understand this effect in the following
way. Andreev levels belong to a many-body system of superconducting
electrons. Although the macroscopic behaviour of the ALQ can be expressed in
terms of the single-particle density matrix, the microscopic interaction with
phonons involves two-particle correlation functions, which are sensitive to
the fermionic nature of the Andreev states and obey the Pauli exclusion
principle. This leads to the reduced probability of the phonon-induced
interlevel transitions and hence to a slower decoherence.

Furthermore, the rate of the phonon-induced transitions between the Andreev
levels is significantly reduced compared to the bulk transition rate. The
reason is that both the Andreev levels belong to the same normal electronic
mode; this together with a rapid spreading out of the Andreev level wave
function in the contact electrode strongly reduces the relevant phonon phase
space.

In the paper, only the case of a single-mode QPC was considered for clarity.
However, the approach might be also relevant for macroscopic Josephson qubits
with tunnel junctions. In junctions with disordered tunnel barriers, the open
conducting modes with large transmissivity are
present.\cite{Bauer97,Likharev00} This introduces low-energy Andreev levels,
which implies that quantum phase fluctuations become coupled to these Andreev
levels, and the system must be described with the effective ALQ-type
Hamiltonian. Another kind of qubits where the effective ALQ Hamiltonian might
be appropriate are d-wave qubits: for the most interesting junction
geometries, low-energy Andreev levels, midgap states, build up in the
junction, which also makes the effective ALQ Hamiltonian to be relevant.

\section{Acknowledgement}

We acknowledge useful discussions on various stages of our work with D.
Khmelnitsky, Yu. Galperin, G. Johansson, A. Zaikin, L. Gorelik, R. Shekhter,
I. Krive, and E. Bezuglyi. Support from EU-SQUBIT consortium, Swedish Science
Foundation, SSF-OXIDE consortium, and KVA is gratefully acknowledged.

% % % % % % % % % % % % APPENDIX % % % % %
\begin{appendix}

\section{Tunnel limit for QPC }

In this Appendix, we consider a low transparency QPC, $D \ll 1$, and connect
our method to the results of Refs. \onlinecite{Ambegaokar,Eckern} for tunnel
Josephson junctions. In this limit, the Andreev levels lie very close to the
edge of the superconducting gap, $E_a \approx \Delta$, and therefore the
field $\eta$ is a fast variable, which has to be integrated out along with
all the other electronic degrees of freedom. This will result in the
effective action for the phase difference alone. After the integration over
electronic fields, the propagator in Eq. (\ref{Ue}) will take the form,
\begin{equation}
{\cal U}_e[\phi] =
\int {\cal D}^2 \eta \, {\cal D}^2 \psi_{L,R} \,
e^{i S_e/\hbar} =
\exp \left\{ i S[\phi]/\hbar \right\} \,,
 %%\label{}
\end{equation}
\begin{equation}
S[\phi]/\hbar = - i {\rm Tr} \, \ln \left( 1 + \frac{D}{(1 + \sqrt{R})^2} \,
g^{-1} \, e^{-i \sigma_z \phi/2} \, g \, e^{i \sigma_z \phi/2} \right) \,.
\label{S:appA}
\end{equation}
Here $g$ is the Green function defined in Eq. (\ref{contact:g}),
\begin{equation}
g(t) = - (1/ \hbar v) \int_{-\infty}^{+\infty} \frac{d \omega}{2 \pi} \, e^{-i
\omega t} \, \frac{\hbar \omega + \Delta \, \sigma_x} {\sqrt{\Delta^2 -
\hbar^2 (\omega + i \, {\rm sgn} \omega \, 0)^2}} \,,
%%\label{}
\end{equation}
and the matrix product in Eq. (\ref{S:appA}) includes also the time
convolutions. Taking advantage of the small $D$, and expanding the action
(\ref{S:appA}), the lowest order term reads:
\begin{equation}
S[\phi]/\hbar = - i (D/4) \, {\rm tr} \int_{-\infty}^{+ \infty} dt_1 dt_2 \,
g^{-1}(t_1 - t_2) \, e^{-i \sigma_z \phi(t_2)/2} g(t_2 -t_1) \, e^{i \sigma_z
\phi(t_1)/2} \,, \label{S2:appA}
\end{equation}
where the trace refers to the Nambu space. After taking the trace, the action
can be written on the following form,
\begin{equation}
S[\phi]/\hbar = \int_{-\infty}^{+ \infty} dt_1 dt_2 \, \left[ \alpha(t_1 -
t_2) \cos \frac{\phi(t_1) - \phi(t_2)}{2} + \beta(t_1 - t_2) \cos
\frac{\phi(t_1) + \phi(t_2)}{2} \right] \,, \label{Seff:appA}
\end{equation}
with the kernels $\alpha$ and $\beta$ given by
\begin{equation}
\alpha(t) = i (D/2) \left( \Delta / 2 \hbar \right)^2 \, \left[ H_1^{(1)}(t
\Delta/\hbar)\right]^2 \,, \label{A:appA}
\end{equation}
\begin{equation}
\beta(t) = - i (D/2) \left( \Delta / 2 \hbar \right)^2 \, \left[ H_0^{(1)}(t
\Delta/\hbar)\right]^2 \,,
\label{B:appA}
\end{equation}
where $H^{(1)}_{0,1}$ are the Hankel functions of the first kind. Analytical
continuation to the imaginary time in Eqs. (\ref{A:appA}) and (\ref{B:appA})
using relation, $K_\nu(t) = (\pi/2) i^{1 + \nu} H_\nu^{(1)}(i t)$, leads to
the same expressions for the kernels, $\alpha$ and $\beta$, as derived in
Refs. \onlinecite{Ambegaokar,Eckern,LarkinOvch83}, for tunnel Josephson
junctions [with normal resistance of the tunnel junction being replaced with
the normal resistance of the single-channel QPC: $R_N = (D e^2/\pi
\hbar)^{-1}$].

For a slow varying phase, $\phi$, on the time scale of the kernel variations,
$\hbar/\Delta$, the both cosine terms in Eq. (\ref{Seff:appA}) can be
expanded over the relative time coordinate, $\tau = t_1-t_2$, up to the
second order, and the effective action takes a simpler form,
\begin{equation}
S[\phi] = \int dt \, \left[ \frac{\delta C}{2} \left( \frac{\hbar
\dot{\phi}}{2 e} \right)^2 + {\hbar\over 2e} I_c \cos \phi \right] \,,
%%\label{}
\end{equation}
where  $I_c = 2 e \int_{-\infty}^{+ \infty} d\tau \, \beta(\tau) = e D \Delta
/2 \hbar$ is the critical Josephson current of the single-channel tunnel
point contact, and
\begin{equation}
\delta C(\phi) = - \frac{e^2}{\hbar} \int_{-\infty}^{+ \infty} d\tau \,
\tau^2 \left[ \, \alpha(\tau) - \beta(\tau) \cos \phi \, \right] =
\frac{3}{32} \frac{D e^2}{\Delta} \left[ \, 1 - (1/3) \cos \phi \, \right]
%%\label{}
\end{equation}
is the correction to the contact capacitance due to the quasiparticle
tunneling.\cite{LarkinOvch83,Eckern}

%%%%%%%%%%%%%%%%%%%%%%%%%%%%%%%%%%%%%%%%%%%%%%%%%%%%%%%%%%%%%%
\section{Derivation of kinetic equation}

In this Appendix we derive Eqs. (\ref{F1}), (\ref{F12}) from the equation for
the Keldysh function, Eq. (\ref{GK_eq_time}).  It is convenient to write the
equation in the mixed representation,
\begin{equation}
i \hbar \partial_t G^K_\omega(t) - E_a^\ast
\left[ \, \sigma_z ,  G^K_\omega(t) \, \right] =
\left( \,
\Sigma^R \, G^K - G^K \, \Sigma^A +
\Sigma^K \, G^A - G^R \, \Sigma^K
\, \right)_\omega (t) \,,
\label{GK_eq}
\end{equation}
where
\begin{equation}
\check{G}_\omega(t) = \int d \tau \, e^{i \omega \tau} \,
\check{G}(t +
\tau/2, t- \tau/2) \,,
%%\label{}
\end{equation}
and the products include the convolutions defined by the equation,
\begin{equation}
\left( \, A \, B \, \right)_\omega(t) =
\int \frac{d \omega_1 d \omega_2}{(2 \pi)^2} \int d\tau_1 d\tau_2 \,
e^{i \left[ (\omega - \omega_1) \tau_2 + (\omega - \omega_2) \tau_1 \right]} \,
A_{\omega_1}(t + \tau_1/2) \, B_{\omega_2}(t - \tau_2/2)
\,.
%%\label{}
\end{equation}
Neglecting effect of the phonons on the
retarded and advanced Green functions of the qubit, $G^{R,A}$, we
replace them in the right hand side of Eq. (\ref{GK_eq}) by the free Green
functions, $G^{R,A}_\omega = \left( \hbar \omega - E_a^\ast \sigma_z \pm i 0
\right)^{- 1}$, which are time-independent in the mixed representation.
Thus, the time-dependence in the self-energy
comes only from the Keldysh component, $G^K$.
The self-energy components take the form,
\begin{equation}
\Sigma^{R(A)}_\omega(t) = (i \hbar /2) \, \sigma_x
\int \frac{d \Omega}{2 \pi} \,
\left[ \,
D^K_{\Omega} \, G^{R(A)}_{\omega - \Omega} +
D^{R(A)}_{\Omega} \, G^K_{\omega - \Omega}(t)
\, \right] \sigma_x
\,,
%%\label{}
\end{equation}
\begin{equation}
\Sigma^K_\omega(t) = (i \hbar /2) \, \sigma_x
\int \frac{d \Omega}{2 \pi} \,
\left( \, D_{\Omega}^R - D_{\Omega}^A \, \right)
\left[ \,
(2 N_{\Omega} + 1) \, G^K_{\omega - \Omega}(t) +
G^R_{\omega - \Omega} - G^A_{\omega - \Omega}
\, \right] \sigma_x
\,,
%%\label{}
\end{equation}
where
\begin{equation}
D_{\Omega}^{R,A} = \int \frac{d \Omega'}{2 \pi \hbar} \,
\frac{{\cal D}_{\Omega'}}{\Omega - \Omega' \pm i 0}
\;,\;\;\;
D_{\Omega}^K = - (i / \hbar) (2 N_\Omega + 1) \, {\cal D}_\Omega
\,,
%%\label{}
\end{equation}
$N_\Omega = \left( e^{\beta \hbar \Omega} - 1 \right)^{-1}$
is the equilibrium phonon distribution function, and
${\cal D}_\Omega$ is the spectral weight function of the phonon bath,
\begin{equation}
{\cal D}_\Omega =
2 \pi \, {\rm sgn}(\Omega) \sum_{\bf q} \, |\gamma_{\bf q}^{*}|^2 \,
\delta(|\Omega| - \Omega_q)
\,.
%%\label{}
\end{equation}
After integrating Eq. (\ref{GK_eq}) over $\omega$, we obtain equation
for the Keldysh function at coinciding times,
\begin{equation}
G^K(t) = \int {d \omega \over 2 \pi} \, G_\omega^K(t),
\label{}
\end{equation}
\begin{equation}
i \hbar \partial_t G^K(t) - E_a^\ast \left[ \, \sigma_z ,  G^K(t) \, \right] =
I_0 + I_1(t) + I_2(t)
\,,
\label{GKt_eq}
\end{equation}
where
$I_0 = (1 / 2 \hbar^2) \, {\cal D}_{2 E_a^\ast/\hbar} \sigma_z$, and
\begin{equation}
I_1(t) = {-i \over 2 \hbar}
\int \frac{d \Omega d \omega}{(2 \pi)^2} \int_0^{+\infty} d \tau \,
(2 N_\Omega + 1) {\cal D}_\Omega  \left[
e^{i (\omega - \Omega + E_a^\ast \sigma_z / \hbar) \tau} \,
G^K_\omega(t - \tau/2) -
e^{i (\omega + \Omega - E_a^\ast \sigma_z / \hbar) \tau} \,
\sigma_x G^K_\omega(t - \tau/2) \sigma_x - h.c. \,
\right],
\label{I1}
\end{equation}
\begin{equation}
I_2(t) =
\int \frac{d \Omega d \omega_1 d \omega_2}{2 (2 \pi)^3}
\int_0^{+\infty} d \tau \,
{\cal D}_\Omega \left[
e^{i (\omega_2 - \omega_1 - \Omega) \tau} \,
\sigma_x G^K_{\omega_1}(t - \tau/2) \sigma_x \,
G^K_{\omega_2}(t - \tau/2) + h.c. \,
\right]
\,.
\label{I2}
\end{equation}

For a weak electron-phonon interaction, the right hand side of Eq.
(\ref{GKt_eq}) is a small perturbation, which allows one to construct an
asymptotic solution by using an improved perturbation expansion,
\begin{equation}
G^K_\omega(t) = - ( 2 \pi i / \hbar) \, e^{-i E_a^\ast \sigma_z t / \hbar}
\left(
F_\omega(\lambda t) +
\sum_{n=1}^{\infty} \lambda^n \tilde{F}^{(n)}_\omega(t)
\right) \, e^{i E_a^\ast \sigma_z t / \hbar}
\,,
\label{ansatz}
\end{equation}
\begin{equation}
F_\omega(t) = \delta(\omega - E_a^\ast \sigma_z / \hbar)
\left(
\begin{array}{cc}
F_1(t) & 0 \\
0 & F_2(t)
\end{array}
\right) + \delta(\omega)
\left(
\begin{array}{cc}
0 & F_{12}(t) \\
F_{12}^\ast(t) & 0
\end{array}
\right)
\,.
\label{F_omegat}
\end{equation}
In Eq. (\ref{ansatz}), $\lambda$ is a formal perturbation parameter,
which reflects weak electron-phonon interaction, $\sim {\cal D}_\Omega$,
and allows one to develop a systematic perturbative expansion. In the
zero-order approximation with respect to $\lambda$, the (time-independent)
matrix $F$ in Eq. (\ref{F_omegat}) is the solution of Eq. (\ref{GKt_eq})
without the right hand side. The first-order equation determines the time
dependence in this matrix (which absorbs formally diverging with time terms
in a straightforward perturbative expansion),
\begin{equation}
\partial_t F_1 = - \partial_t F_2 =
- {\nu \over 2} \left[ \, (2N+1) \left( F_1 - F_2 \right) +  F_1 F_2 -
1  \, \right] \,,
\label{F1ap}
\end{equation}
\begin{equation}
\partial_t F_{12} = - \left[ \, { \nu  \over 2 } \, \left(
2 N + 1 - F_z \right) + 2 i (\delta + \delta_0 F_z) \, \right] F_{12}
\;,\;\;\; F_z(t) = \left( F_1 - F_2 \right)/2 \,,
\label{F12ap}
\end{equation}
where  $\nu =  {\cal D}_{2 E_a^\ast/\hbar } \, / \hbar^2$
is the phonon-induced transition rate between the qubit levels,
$N = N_{2 E_a^\ast / \hbar}$, and quantities
\begin{equation}
\delta = (1 / 2 \hbar^2)
{\int\;\!\!\!\!\!\!\!\!- \;}
\frac{d \Omega}{2 \pi} \, (2 N_\Omega + 1)
\frac{{\cal D}_\Omega }{\Omega + 2 E_a^\ast / \hbar} \;,\;\;\;
\delta_0 = (1 / 2 \hbar^2)
{\int\;\!\!\!\!\!\!\!\!- \;}
\frac{d \Omega}{2 \pi} \, \frac{{\cal D}_\Omega }{\Omega + 2 E_a^\ast / \hbar}
\,,
\label{deltas}
\end{equation}
determine the phonon-induced shift of the qubit frequency. The higher order
equations determine the rapidly oscillating terms, $\tilde{F}^{(n)}_\omega$,
in Eq. (\ref{ansatz}); e.g., equation for $\tilde{F}^{(1)}_\omega$ reads,
\begin{equation}
\partial_t \tilde{F}^{(1)} =
{\cal I}_z \sigma_z + {\cal I}_{+} \sigma_{+} + {\cal I}_{-} \sigma_{-}
\,,
%%\label{}
\end{equation}
where $\sigma_\pm = (1/2) \left( \sigma_x \pm i \sigma_y \right)$,
\begin{equation}
{\cal I}_z  = e^{-4 i E_a^\ast t/ \hbar} \left(
\nu/4 - i \, \delta_0
\right) F_{12}^2(t) + c.c. \,,
%%\label{}
\end{equation}
\begin{equation}
{\cal I}_{-} = {\cal I}_{+}^\ast =
e^{- 4 i E_a^\ast t/\hbar} \left[ \,
\nu \, (N + 1/2) + 2 i \left( \delta  + \tilde{\delta}_0 \,  F_z \right)
\, \right] F_{12}(t)
\;,\;\;\;
\tilde{\delta}_0 = (1 / 2 \hbar^2)
{\int\;\!\!\!\!\!\!\!\!- \;}
\frac{d \Omega}{2 \pi} \, \frac{{\cal D}_\Omega }{\Omega}
\,.
%%\label{}
\end{equation}
It follows from these equations that $\tilde{F}^{(1)}$ indeed rapidly
oscillates, with the frequency, $4 E_a^\ast/\hbar$, and has relatively small
amplitude, proportional to $\hbar \nu / E_a^\ast$.

% % % % % % % % % % % % % % % % % % % % % % % % % % % % % % % % % % % % % %

\end{appendix}

% % % % % % % % % BIBLIOGRAPHY % % % % % % % % % % % % % % % % % % % % % % % %

\end{document}